\documentclass[aps,pra,showpacs,superscriptaddress,amssymb,twocolumn]{revtex4-1}
\usepackage{graphicx}
\usepackage{appendix} 
\usepackage{hyperref}
\usepackage{bm,amsmath}
\usepackage{dcolumn}

\begin{document}


\vspace{0.2in}

\title{Tunable coupler for superconducting Xmon qubits: Perturbative nonlinear model}

\author{Michael R. Geller}
\email{mgeller@uga.edu}
\affiliation{Department of Physics and Astronomy, University of Georgia, Athens, Georgia 30602, USA}

\author{Emmanuel Donate}
\affiliation{Department of Physics and Astronomy, University of Georgia, Athens, Georgia 30602, USA}

\author{Yu Chen}
\affiliation{Department of Physics, University of California, Santa Barbara, California 93106, USA}

\author{Charles Neill}
\affiliation{Department of Physics, University of California, Santa Barbara, California 93106, USA}

\author{Pedram Roushan}
\affiliation{Department of Physics, University of California, Santa Barbara, California 93106, USA}

\author{John M. Martinis}
\affiliation{Department of Physics, University of California, Santa Barbara, California 93106, USA}

\date{\today}

\begin{abstract}

We study a recently demonstrated design for a high-performance tunable coupler suitable for superconducting Xmon and planar transmon qubits [Y.~Chen {\it et al.}, arXiv:1402.7367]. The coupler circuit uses a single flux-biased Josephson junction and acts as a tunable current divider. We calculate the effective qubit-qubit interaction Hamiltonian by treating the nonlinearity of the qubit and coupler junctions perturbatively. We find that the qubit nonlinearity has two principal effects: The first is to suppress the magnitude of the transverse $\sigma^x \! \otimes \sigma^x$ coupling from that obtained in the harmonic approximation by about 15\%. The second is to induce a small diagonal $\sigma^z \! \otimes \sigma^z$ coupling. The effects of the coupler junction nonlinearity are negligible in the parameter regime considered. 
\end{abstract}


\maketitle

\section{Introduction}
\label{introduction section}

The development of a fully planar transmon-type superconducting qubit, which combines high coherence with several other features desirable for logic gate implementation and scalability, could make a quantum computer  based on quantum integrated circuits possible in the near future \cite{BarendsPRL13}. These Xmon qubits can be directly wired together (or to a resonator bus) with fixed capacitors \cite{BarendsNat14}, but the resulting couplings are always present and degrade gate performance. A simple tunable coupler option is therefore desirable. Tunable coupling is also required for the single-excitation subspace method of Ref.~\cite{GellerMartinisSornborgerEtalPre12}, and may also be desirable for analog quantum simulation applications.  Although  a wide variety of tunable coupler designs for superconducting circuits have been considered previously \cite{MakhlinNat99,BlaisPRL03,PlourdePRB04,LantzPRB04,GrajcarPRB06,HimeSci06,vanderPloegPRL07,NiskanenSci07,HarrisPRL07,AshhabPRB08,YamamotoPRB08,MariantoniPRB08,AllmanPRL10,PintoPRB10,GambettaPRL11,BialczakPRL11,SrinivasanPRL11,GroszkowskiPRB11,HoffmanPRB11}, most of these designs are intended for flux qubits. The coupler we discuss in this work is suitable for Xmon \cite{BarendsPRL13} and planar transmon \cite{KochPRA07} qubits, which have no trapped flux, and the design is experimentally practical. Most importantly, the design introduces tunability without compromising high coherence. Tunably coupled Xmon's based on this design, which are called {\it gmon} qubits, have been demonstrated recently \cite{ChenEtal1402.7367}.

\begin{figure}
\includegraphics[width=8.0cm]{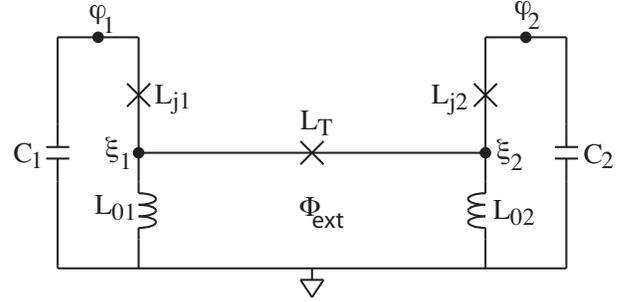} 
\caption{Coupler circuit schematic. The $\varphi_i$ and $\xi_i$ are node flux variables, and $\Phi_{\rm ext}$ is an external magnetic flux bias. There are four active nodes (black dots) in this circuit. The Josephson junctions labelled $L_{\rm j}$ are each double junctions threaded by additional fluxes (not shown) that tune the qubit frequencies.}
\label{coupler circuit figure}
\end{figure}

The coupler circuit is shown in Fig.~\ref{coupler circuit figure}. We begin by discussing the circuit in the harmonic approximation. Josephson junctions (crosses) are characterized by their zero-bias linear inductances $L_{\rm j}$ and $L_{\rm T}$. In particular, $L_{\rm T} = \Phi_0/2 \pi I_{\rm c}$, where $\Phi_0 \!  \equiv \!  h/2e$ and $I_{\rm c}$ is the critical current of the coupler junction. A magnetic flux bias $\Phi_{\rm ext}$ is used to tune coupler junction's effective linear inductance to 
\begin{equation}
L_{\rm eff} = \frac{L_{\rm T}}{\cos \delta} \, ,
\label{Leff definition}
\end{equation}
where $\delta$ is the DC phase difference across the coupler. The relation between $\delta$ and $\Phi_{\rm ext}$ follows from writing the total magnetic flux $\Phi  \equiv \oint_{\Gamma} {\bf A} \cdot d{\bf l} = (\delta/2\pi)\Phi_0$ in the coupler loop $\Gamma$ as 
\begin{equation}
\Phi  = \Phi_{\rm ext} - L_{\rm loop} I_{\rm c} \sin \delta,
\label{flux screening equation}
\end{equation}
where $L_{\rm loop} = L_{01} + L_{02}$ (no DC current flows through the capacitors). Here $I_{\rm c} \sin \delta$ is the induced supercurrent. Then (\ref{flux screening equation}) leads to
\begin{equation}
\delta + \bigg(\frac{L_{01}+L_{02}}{L_{\rm T}}\bigg) \sin\delta = \phi_{\rm ext},
\label{delta equation}
\end{equation}
where
\begin{equation}
\phi_{\rm ext} \! \equiv \! 2 \pi \frac{\Phi_{\rm ext}}{\Phi_0}.
\end{equation}

When $L_{\rm eff} \! \rightarrow \! \infty$, no AC current flows through the coupler and the circuit describes two uncoupled qubits. This occurs when 
\begin{equation}
\delta \ {\rm mod} \ 2\pi = \bigg( \frac{\pi}{2} \, , \frac{3\pi}{2} \bigg).
\end{equation}
Then (\ref{delta equation}) shows that the coupling vanishes when
\begin{equation}
\phi_{\rm ext} \ {\rm mod} \ 2\pi = \bigg( \frac{\pi}{2} + \frac{L_{01}+L_{02}}{L_{\rm T}} \, , \frac{3\pi}{2} - \frac{L_{01}+L_{02}}{L_{\rm T}} \bigg).
\label{coupling zeros}
\end{equation}
In the weakly coupled limit the effective coupling strength---half the splitting between the symmetric and antisymmetric eigenstates---is approximately \cite{ChenEtal1402.7367}
\begin{equation}
g = -  \frac{ L_0^2 \cos \delta} {2(L_{\rm j} + L_0)(L_{\rm T} + 2 L_0 \cos \delta) } \, \omega_{\rm q},
\label{simplified coupling}
\end{equation}
where $\omega_{\rm q}$ is the qubit frequency. In (\ref{simplified coupling}) we have assumed identical qubits in resonance. 

\begin{figure}
\includegraphics[width=6.0cm]{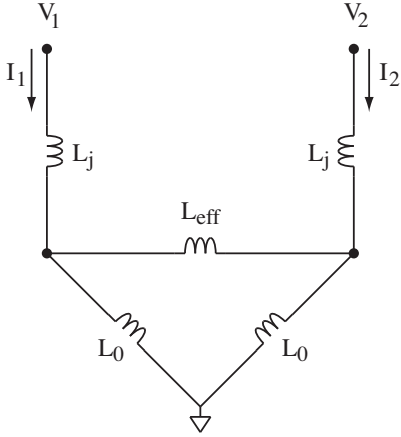} 
\caption{Network of linear inductors.}
\label{inductive network figure}
\end{figure} 

The expression (\ref{simplified coupling}) 
is valid in the weak coupling limit and, in addition, does not account for qubit and coupler anharmonicity (beyond the flux-dependence of the linear inductance $L_{\rm eff}$). It can be derived, essentially classically, from the input impedances to the network of Fig.~\ref{inductive network figure}, defined through the relation
\begin{equation}
\begin{pmatrix}
V_1  \\
V_2  \\
\end{pmatrix}
= 
\begin{pmatrix}
L_{\rm q} & M  \\
M & L_{\rm q}  \\
\end{pmatrix}
\begin{pmatrix}
{\dot I}_1  \\
{\dot I}_2 \\
\end{pmatrix}.
\label{impedance equation}
\end{equation}
We find
\begin{equation}
M = \frac{L_0^2}{L_{\rm eff} + 2 L_0}
\ \ \ {\rm and} \ \ \ 
L_{\rm q} = L_{\rm j} + L_0 - M.
\label{input impedances}
\end{equation}
The potential energy of the circuit in Fig.~\ref{coupler circuit figure} in the harmonic approximation is therefore
\begin{equation}
U = \bigg(\frac{\Phi_0}{2\pi}\bigg)^{\! \! 2}
\bigg[ \frac{\varphi_1^2}{2KL_{\rm q}} +  \frac{\varphi_2^2}{2KL_{\rm q}}
+ \Gamma_{11} \varphi_1 \varphi_2 \bigg],
\label{potential energy with cross term} 
\end{equation}
where
\begin{equation}
K = 1- \bigg(\frac{M}{L_{\rm q}} \bigg)^2 
\ \ \ {\rm and} \ \ \
\Gamma_{11} = - \frac{M}{K L_{\rm q}^2}.
\label{K and Gamma11 definitions}
\end{equation}
In the weakly coupled limit, $M \ll L_{\rm q}$. To obtain  (\ref{simplified coupling}) we assume 
\begin{eqnarray}
&K \approx 1,& \\
\label{K in weak coupling}
&L_{\rm q}  \approx  L_{\rm j} + L_0,&
\label{Lq in weak coupling}
\end{eqnarray}
and
\begin{equation}
\Gamma_{11}  \approx 
-\frac{L_0^2}{(L_{\rm j}+L_0)^2(L_{\rm eff}+ 2 L_0)}.
\label{Gamma11 in weak coupling}
\end{equation}
These approximations will be removed in Sec.~\ref{perturbative nonlinear section}.

Next we calculate the splitting induced by (\ref{Gamma11 in weak coupling}). We can again compute this classically by treating the qubits as LC oscillators with frequency $\omega_{\rm q}=(L_{\rm q}C)^{-\frac{1}{2}}$, where $L_{\rm q}$ is given by (\ref{Lq in weak coupling}). The potential energy of the coupled oscillators is given in (\ref{potential energy with cross term}) with $K=1$. Diagonalizing the quadratic form (\ref{potential energy with cross term}) leads to eigenmodes with shifted inductances $1/(L_{\rm q}^{-1} \pm \Gamma_{11})$ and hence frequencies $\sqrt{1 \pm L_{\rm q} \Gamma_{11} } \, \omega_{\rm q}$. Therefore in the weakly coupled limit we obtain
\begin{equation}
g =  \frac{\Gamma_{11} L_{\rm q}}{2} \, \omega_{\rm q},
\label{simple g}
\end{equation} 
a result that also applies to coupled qubits (see below) and leads to (\ref{simplified coupling}). 

\begin{table}[htb]
\centering
\caption{\label{parameter table} Example values of circuit parameters.}
\begin{tabular}{|c|c|}
\hline
quantity  & value \\
\hline 
$C_1, C_2$  & $91 \, {\rm fF}$  \\
\hline 
$L_{{\rm j}1}, L_{{\rm j}2}$  & $8.6 \, {\rm nH}$  \\
\hline 
$L_{01}, L_{02}$  & $200 \, {\rm pH}$  \\
\hline 
$L_{\rm T}$  & $1.3 \, {\rm nH}$  \\
\hline 
\end{tabular}
\end{table}

\begin{figure}
\includegraphics[width=9cm]{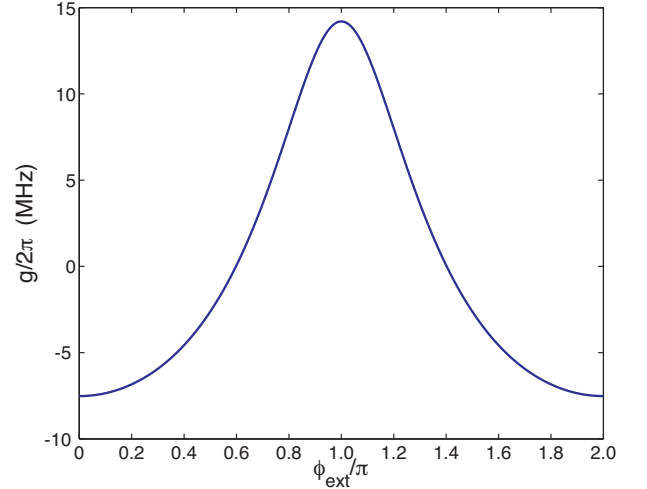} 
\caption{(Color online) Coupling strength in the weak coupling limit (\ref{simplified coupling}), for system parameters given in Table~\ref{parameter table}.}
\label{approximate coupling figure}
\end{figure} 

In Table~\ref{parameter table} we provide an example of possible system parameter values. The approximate coupling function (\ref{simplified coupling}) for these parameters is shown in Fig.~\ref{approximate coupling figure}. Here $\delta(\phi_{\rm ext})$ is obtained from (\ref{delta equation}). With these parameter values the coupling to vanishes at [see (\ref{coupling zeros})] 
\begin{equation}
\phi_{\rm ext} \ {\rm mod} \ 2\pi = \big( 0.598\pi , 1.402\pi \big).
\label{coupling zeros evaluated}
\end{equation}
In the remainder of this paper we calculate the transverse $\sigma^x \otimes \sigma^x$ coupling $g$, going beyond the approximations leading to (\ref{simplified coupling}), and we also compute the diagonal $\sigma^z \otimes \sigma^z$ coupling. In Sec.~\ref{nonlinear circuit model section} we construct the Hamiltonian for the nonlinear circuit and calculate the transverse coupling numerically by exact diagonalization. In Sec.~\ref{perturbative nonlinear section} we calculate the coupling for the linearized model beyond weak coupling and study the nonlinearity perturbatively. In Sec.~\ref{diagonal coupling section} we calculate the $\sigma^z \otimes \sigma^z$ coupling, both analytically and numerically.

\section{NONLINEAR CIRCUIT MODEL}
\label{nonlinear circuit model section}

The state of the circuit in Fig.~\ref{coupler circuit figure} is described by four coordinates. However the $\xi_1$ and $\xi_2$ nodes have negligible capacitance to ground, and therefore are ``massless" degrees of freedom that remain in their instantaneous ground states. They will be eliminated from the problem in the analysis below. The complete Lagrangian for the circuit of Fig.~\ref{coupler circuit figure} is
\begin{equation}
L= \sum_{i=1,2}  \bigg(\frac{\Phi_0}{2\pi}\bigg)^{\! \! 2}  
\frac{C_i}{2} {\dot \varphi_i}^2 \ - \ U,
\label{circuit lagrangian}
\end{equation}
where
\begin{eqnarray}
U &=& \sum_{i=1,2} \bigg\lbrace \bigg(\frac{\Phi_0}{2\pi}\bigg)^{\! \! 2}  
\bigg[ \frac{\xi_i^2 }{2 L_{0i}} 
- \frac{\cos(\varphi_i-\xi_i )}{L_{{\rm j}i}}  \bigg]
\bigg\rbrace \nonumber \\
&-&  \bigg( \frac{\Phi_0}{2\pi}\bigg)^{\! \! 2}
\frac{\cos(\xi_1 - \xi_2 - \phi_{\rm ext})}{L_{\rm T}} .
\label{circuit potential}
\end{eqnarray}

\begin{figure}
\includegraphics[width=9.0cm]{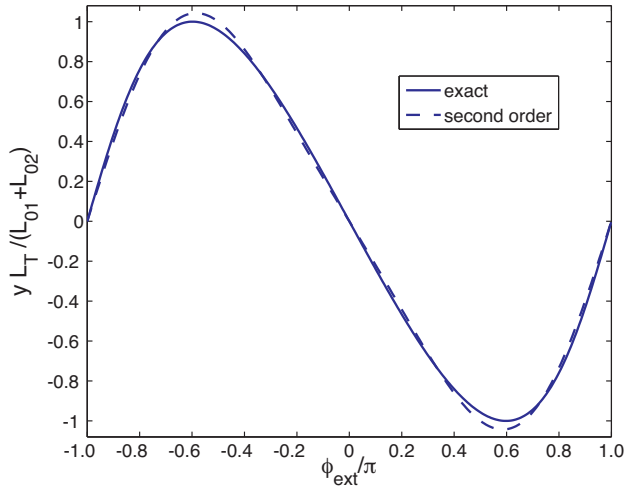} 
\caption{(Color online) Exact and second-order approximate solution of equilibrium condition (\ref{y equation}).}
\label{y figure}
\end{figure} 

We begin our analysis by writing the four coordinates as classical equilibrium or DC values that minimize the potential energy (\ref{circuit potential}), plus deviations. Two of the four equilibrium conditions lead to 
\begin{equation}
{\bar \varphi}_i = {\bar \xi}_i,
\label{equilibrium conditions 1&2}
\end{equation}
where the bar denotes equilibrium values. The remaining two conditions can be written as
\begin{equation}
\frac{ {\bar \xi}_1}{L_{01}} = -x
\ \ \ {\rm and} \ \ \
\frac{ {\bar \xi}_2}{L_{02}} = x,
\label{equilibrium conditions 3&4}
\end{equation}
where
\begin{equation}
x \equiv \frac{\sin({\bar \xi}_1- {\bar \xi}_2 - \phi_{\rm ext} )}{L_{\rm T}}.
\label{y definition}
\end{equation}
Combining (\ref{equilibrium conditions 3&4}) and (\ref{y definition}) leads to 
\begin{equation}
x = - \frac{\sin[ (L_{01} + L_{02}) \, x + \phi_{\rm ext} ]}{L_{\rm T}}.
\label{x equation}
\end{equation}
We solve ({\ref{x equation}) approximately, in the weak coupling limit. To do this we define $y \equiv (L_{01} + L_{02}) \, x,$ which leads to
\begin{equation}
y = - \frac{L_{01}+L_{02}}{L_{\rm T}} \sin( y + \phi_{\rm ext} ).
\label{y equation}
\end{equation}
Solving (\ref{y equation}) iteratively leads to a solution expressed as a power series in $(L_{01}+L_{02})/L_{\rm T}$. The solution to second order is
\begin{equation}
y = - \frac{L_{01}+L_{02}}{L_{\rm T}} \sin (\phi_{\rm ext}) +  \frac{1}{2} \bigg(\frac{L_{01}+L_{02}}{L_{\rm T}}\bigg)^2 \! \! \sin (2 \phi_{\rm ext}),
\label{approximate y solution}
\end{equation}
which is plotted in Fig.~\ref{y figure} along with the exact numerical solution for the parameters in Table~\ref{parameter table}. Putting everything together we obtain
\begin{equation}
{\bar \varphi}_1 = {\bar \xi}_1 = \frac{L_{01}}{L_{\rm T}} \bigg( \! \sin \phi_{\rm ext} - \frac{L_{01} + L_{02}}{2L_{\rm T}} \, \sin 2\phi_{\rm ext} \bigg)
\label{xi1 bar perturbative result}
\end{equation}
and
\begin{equation}
{\bar \varphi}_2 = {\bar \xi}_2 = - \frac{L_{02}}{L_{\rm T}} \bigg(  \! \sin \phi_{\rm ext} - \frac{L_{01} + L_{02}}{2L_{\rm T}} \,\sin 2\phi_{\rm ext} \bigg).  
\label{xi2 bar perturbative result}
\end{equation}

\begin{figure}
\includegraphics[width=9.0cm]{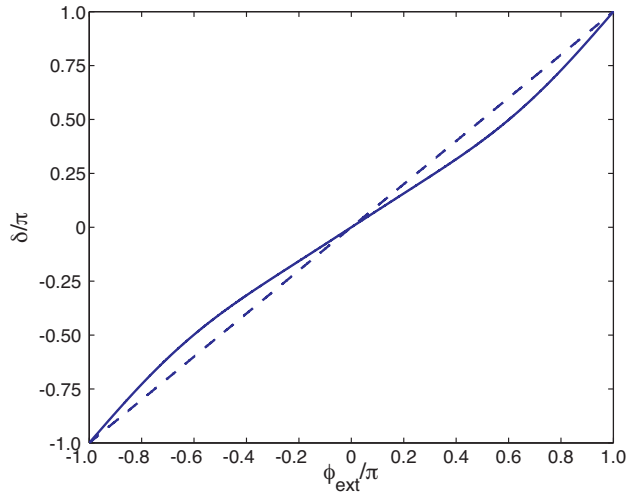} 
\caption{(Color online) Plot of function $\delta$, defined in (\ref{delta vs flux}), assuming circuit parameter values given in Table \ref{parameter table}. The dashed line is the function $\delta=\phi_{\rm ext}$.}
\label{f figure}
\end{figure} 

Finally, we rewrite the circuit Lagrangian (\ref{circuit lagrangian}) and (\ref{circuit potential}) in terms of the equilibrium coordinates. After a change of variables
\begin{eqnarray}
\varphi_i &\rightarrow& {\bar \varphi}_i + \varphi_i , \nonumber \\
\xi_i & \rightarrow& {\bar \xi}_i + \xi_i , 
\end{eqnarray}
the potential (\ref{circuit potential}) becomes 
\begin{eqnarray}
U &=& \sum_{i=1,2} \bigg\lbrace \bigg(\frac{\Phi_0}{2\pi}\bigg)^{\! \! 2}  \bigg[
\frac{({\bar \xi}_i + \xi_i)^2 }{2 L_{0i}} 
- \frac{\cos(\varphi_i-\xi_i )}{L_{{\rm j}i}}  
\bigg] \bigg\rbrace \nonumber \\
&-& \bigg( \frac{\Phi_0}{2\pi}\bigg)^{\! \! 2}
\frac{\cos(\xi_1 - \xi_2 - \delta)}{L_{\rm T}} ,
\label{circuit potential with equilibrium coordinates}
\end{eqnarray}
where the $\varphi_i$ and $\xi_i$ variables now denote {\it deviations} from equilibrium, and
\begin{equation}
\delta \equiv \phi_{\rm ext} + {\bar \xi}_2 - {\bar \xi}_1
\label{delta vs flux}
\end{equation}
which is plotted in Fig.~\ref{f figure}. The function (\ref{delta vs flux}) relates the DC phase difference $\delta$ across the coupler junction to the external flux. The definition (\ref{delta vs flux}) of $\delta$ is equivalent to that used in (\ref{Leff definition}), and the dependence on $\phi_{\rm ext}$ resulting from (\ref{xi1 bar perturbative result}) and (\ref{xi2 bar perturbative result}) is equivalent to that obtained by solving (\ref{delta equation}) perturbatively in the small $(L_{01}+L_{02})/L_{\rm T}$ limit.
Note that in this limit the approximation
\begin{equation}
\delta \approx \phi_{\rm ext} ,\label{approximate f}
\end{equation}
can sometimes be used, which is also shown in Fig.~\ref{f figure}.

Now we are ready to construct the Hamiltonian: The momentum conjugate to $\varphi_i$ is 
\begin{equation}
p_i = \bigg( \frac{\Phi_0}{2\pi}\bigg)^{\! \! 2} \! C_i \, {\dot \varphi_i}.
\end{equation}
The momenta conjugate to the $\xi_i$ vanish. The complete Hamiltonian for the circuit of Fig.~\ref{coupler circuit figure} is therefore
\begin{equation}
H=  \bigg(\frac{2\pi}{\Phi_0}\bigg)^{\! \! 2}  \sum_{i}
\frac{p_i^2}{2C_i} \ + \ U,
\label{full hamiltonian}
\end{equation}
where $U$ is given in (\ref{circuit potential with equilibrium coordinates}).

In Fig.~\ref{g exact diagonalization figure} we plot the splitting between the symmetric and antisymmetric eigenstates---equal to twice the magnitude of the transverse component of the effective coupling strength---for the full nonlinear model (\ref{circuit potential with equilibrium coordinates}), assuming the circuit parameters given in Table \ref{parameter table}. To obtain these results we use a two-dimensional grid in the coordinates $\varphi_1$ and $\varphi_2$, and the basis 
\begin{equation}
\big| \varphi_1 , \varphi_2 \big\rangle
\ \ {\rm with} \ \ \varphi_i \in 
\lbrace -\pi, -\pi + d\varphi , \dots , 0, \dots , \pi \rbrace,
\label{grid basis}
\end{equation} 
where $d\varphi$ is the mesh spacing. The kinetic energy operator for coordinate $\varphi_1$ is approximated as
\begin{eqnarray}
{\rm KE}_1 \ \big|\varphi_1,\varphi_2\big\rangle =
-  \frac{\hbar^2}{2 C_1 (\Phi_0/2\pi)^2 \, d\phi^2} 
\nonumber \\ 
\times \bigg(  \big|\varphi_1 + d\phi,\varphi_2\big\rangle +  \big|\varphi_1 - d\phi,\varphi_2\big\rangle  \bigg),
\label{tight-binding KE operator}
\end{eqnarray}
and similarly for that of $\varphi_2$. This ``tight-binding" approximation replaces the quadratic kinetic energy in (\ref{full hamiltonian}) by a cosine with the same curvature. We note that the factor of $\hbar^2$ in the numerator of (\ref{tight-binding KE operator}) is required because the $p_i$ in (\ref{full hamiltonian}) are dimensionless. The potential energy is diagonal in the basis (\ref{grid basis}), and for each $| \varphi_1 , \varphi_2 \rangle$ is found by numerically 
minimizing the potential (\ref{circuit potential with equilibrium coordinates}) with respect to the two massless variables $\xi_1$ and $\xi_2$. The exact diagonalization result is shown in the solid curve in Fig.~\ref{g exact diagonalization figure} along with that of the harmonic approximation $2|g|$ and the perturbative result of 
Sec.~\ref{perturbative nonlinear section}.

\begin{figure}
\includegraphics[width=9.0cm]{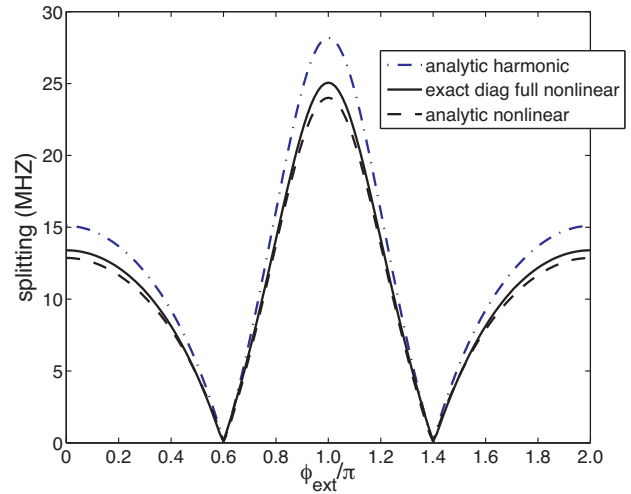} 
\caption{(Color online) Splitting (equal to twice the magnitude of the coupling) in the fully nonlinear model (\ref{full hamiltonian}) calculated by exact diagonalization (solid curve). Also shown are the corresponding harmonic approximation (dashed-dotted curve) and perturbative nonlinear (dashed) results.}
\label{g exact diagonalization figure}
\end{figure} 

\section{PERTURBATIVE TREATMENT OF NONLINEARITY}
\label{perturbative nonlinear section}

In this section we show that the form and strength of the qubit-qubit coupling can be derived analytically by treating the nonlinearity in (\ref{circuit potential with equilibrium coordinates}) perturbatively. First we expand (\ref{circuit potential with equilibrium coordinates}) in powers of the deviations $\varphi_i$ and $\xi_i$, keeping all terms to quartic order. This leads to
\begin{eqnarray}
U &=& \sum_{i=1,2} \bigg\lbrace \bigg(\frac{\Phi_0}{2\pi}\bigg)^{\! \! 2}  \bigg[
\frac{\xi_i^2 }{2 L_{0i}} 
+ \frac{(\varphi_i-\xi_i )^2}{2 L_{{\rm j}i}}  
- \lambda \frac{(\varphi_i-\xi_i )^4}{24 L_{{\rm j}i}}  
\bigg] \bigg\rbrace \nonumber \\
&+&  \bigg( \frac{\Phi_0}{2\pi}\bigg)^{\! \! 2} \bigg[
\cos(\delta) \, \frac{(\xi_1 - \xi_2)^2}{2 L_{\rm T}} 
+ \lambda' \sin(\delta) \,
\frac{(\xi_1 - \xi_2)^3}{6 L_{\rm T}} 
\nonumber \\
&-& \lambda' \cos(\delta)  \,
\frac{(\xi_1 - \xi_2)^4}{24 L_{\rm T}} \bigg]
+ {\rm const.},
\label{nonlinear perturbative potential}
\end{eqnarray}
where parameters $\lambda = 1$ and $\lambda' = 1$ have been introduced to track powers of the qubit and coupler junction nonlinearities, respectively. Note that the first order terms vanish on account of conditions (\ref{equilibrium conditions 1&2}) and (\ref{equilibrium conditions 3&4}), and that the coupler junction induces both cubic and quartic nonlinearity. In this section we develop a theory of the coupling to first order in $\lambda$ and $\lambda^\prime$, neglecting all second order corrections, including those of order $\lambda \lambda^\prime$.

Because there is no kinetic energy associated with the massless $\xi_i$ coordinates, we can eliminate them from the Hamiltonian by replacing 
$U(\varphi_1,\varphi_2,\xi_1,\xi_2)$ with $U(\varphi_1,\varphi_2,\xi_1^*,\xi_2^*)$, where the $\xi^{*}_i$ minimize (\ref{nonlinear perturbative potential}) for fixed $\varphi_i$. This procedure is different that what we did above in (\ref{equilibrium conditions 1&2}) and (\ref{equilibrium conditions 3&4}), because there we minimized $U$ with respect to all four coordinates. Differentiation of (\ref{nonlinear perturbative potential}) with respect to the $\xi_i$ leads to a pair of equations that can be written as
\begin{eqnarray}
\frac{\xi_1^*}{L_{\Sigma 1}}  - \cos(\delta) \, \frac{\xi_2^*}{L_{\rm T}} =
\frac{\varphi_1}{L_{{\rm j}1}} - \lambda \frac{(\varphi_1 - \xi_1^*)^3}{6 L_{{\rm j}1}} \nonumber \\   
- \lambda^\prime \sin(\delta) \, \frac{(\xi_1^* - \xi_2^*)^2}{2 L_{\rm T}}
+ \lambda^\prime \cos(\delta) \, \frac{(\xi_1^* - \xi_2^*)^3}{6 L_{\rm T}}     
\label{simultaneous equation for xi1*}
\end{eqnarray}
and
\begin{eqnarray}
\frac{\xi_2^*}{L_{\Sigma 2}}  - \cos(\delta) \, \frac{\xi_1^*}{L_{\rm T}} =
\frac{\varphi_2}{L_{{\rm j}2}} - \lambda \frac{(\varphi_2 - \xi_2^*)^3}{6 L_{{\rm j}2}} \nonumber \\   
+ \lambda^\prime \sin(\delta) \, \frac{(\xi_1^* - \xi_2^*)^2}{2 L_{\rm T}}
- \lambda^\prime \cos(\delta) \, \frac{(\xi_1^* - \xi_2^*)^3}{6 L_{\rm T}} ,    
\label{simultaneous equation for xi2*}
\end{eqnarray}
where
\begin{equation}
\frac{1}{L_{\Sigma i}} \equiv \frac{1}{L_{{\rm j}i}}
+\frac{1}{L_{0i}} +  \frac{\cos(\delta)}{L_{\rm T}}.
\label{L sum definition}
\end{equation}

We solve the coupled nonlinear equations (\ref{simultaneous equation for xi1*}) and (\ref{simultaneous equation for xi2*}) perturbatively, to first order in $\lambda$ and $\lambda^\prime$, by expanding
\begin{equation}
\xi_i^* = \xi_i^{(0)} + \xi_i^{(1)} \! \! ,
 \ \ \ \ \ (i=1,2)
 \label{xi expansion}
\end{equation}
where the $\xi_i^{(0)}$ are zeroth order in the nonlinearity and the $\xi_i^{(1)}$ are first order. The zeroth order solutions are
\begin{equation}
\xi_i^{(0)} =
\alpha_{i} \varphi_i + \beta_{\bar i} \varphi_{\bar i},
\label{xi zeroth order}
\end{equation}
where
\begin{equation}
\alpha_i \equiv \frac{1}{L_{{\rm j}i} L_{\Sigma{\bar i}}D} ,
\ \ \  \ \ \beta_i \equiv  \frac{\cos(\delta)}{L_{{\rm j}i} L_{\rm T}D}, 
\end{equation}
and where ${\bar i}$ is the index complement to $i$: 
\begin{equation}
{\bar 1} = 2 \ \ \ {\rm and} \ \ \ {\bar 2} = 1.
\end{equation}
Here
\begin{equation}
D \equiv  \frac{1}{L_{\Sigma 1}L_{\Sigma 2}} - \frac{\cos^2(\delta)}{L_{\rm T}^2} .
\end{equation}
The first order corrections are
\begin{eqnarray}
\xi_1^{(1)} &=& - \frac{\lambda}{6} \bigg( \alpha_1 
\big[\varphi_1 - \xi_1^{(0)}\big]^3 + \beta_2 \big[\varphi_2 - \xi_2^{(0)}\big]^3
\bigg) \nonumber \\
&+& \lambda^\prime \, \frac{A}{D} \bigg( \frac{1}{L_{\Sigma 2}} - \frac{\cos(\delta)}{L_{\rm T}} \bigg), \nonumber \\
\xi_2^{(1)} &=& - \frac{\lambda}{6} \bigg( \alpha_2 
\big[\varphi_2 - \xi_2^{(0)}\big]^3 + \beta_1 \big[\varphi_1 - \xi_1^{(0)}\big]^3
\bigg) \nonumber \\
&-& \lambda^\prime \, \frac{A}{D} \bigg( \frac{1}{L_{\Sigma 1}} - \frac{\cos(\delta)}{L_{\rm T}} \bigg),
\label{xi correction}
\end{eqnarray}
where
\begin{eqnarray}
A &\equiv& - \frac{\sin(\delta)}{2 L_{\rm T}} \bigg[ (\alpha_1-\beta_1) \, \varphi_1 - (\alpha_2-\beta_2) \, \varphi_2 \bigg]^2 \nonumber \\
&+& \frac{\cos(\delta)}{6 L_{\rm T}} \bigg[ (\alpha_1-\beta_1) \, \varphi_1 - (\alpha_2-\beta_2) \, \varphi_2 \bigg]^3 \! \! .
\label{A definition}
\end{eqnarray}

Using (\ref{xi zeroth order}) and (\ref{xi correction}) we obtain
\begin{equation}
H =  \sum_{i=1,2} \bigg(\frac{2\pi}{\Phi_0}\bigg)^{\! \! 2} 
\frac{p_i^2}{2C_i} \ + \ U^{(0)} + \ U^{(1)}, 
\label{full hamiltonian expanded potential}
\end{equation}
where
\begin{equation}
U^{(0)}  = \sum_{i=1,2}  \bigg( \frac{\Phi_0}{2\pi}\bigg)^{\! \! 2} \frac{\varphi_i^2}{2 L_{{\rm q}i}}  
+ \bigg( \frac{\Phi_0}{2\pi}\bigg)^{\! \! 2} \, \Gamma_{11} \, \varphi_1 \varphi_2,
\label{U0}
\end{equation}
\begin{equation}
\frac{1}{L_{{\rm q}i}} \equiv 
\frac{(1-\alpha_{i})^2}{L_{{\rm j}i}} 
+ \frac{\alpha_{i}^2}{L_{0i}} 
+ \frac{\beta_{i}^2}{L_{{\rm j} {\bar i}}} 
+ \frac{\beta_{i}^2}{L_{0{\bar i}}} 
+ \cos(\delta) \, \frac{(\alpha_{i}-\beta_{i})^2}{L_{\rm T}}  , 
\label{Lq}
\end{equation}
\begin{eqnarray}
\Gamma_{11} &\equiv&  
\frac{(\alpha_1-1) \beta_2}{L_{\rm j 1}} 
+ \frac{(\alpha_2-1) \beta_1}{L_{\rm j 2}} 
+ \frac{\alpha_1 \beta_2}{L_{01}} 
+ \frac{\alpha_2 \beta_1}{L_{02}} 
\label{gamma11}
\nonumber \\
&-& \cos(\delta) \, \frac{(\alpha_1-\beta_1)(\alpha_2-\beta_2)}{L_{\rm T}} ,
\end{eqnarray}
and where
\begin{eqnarray}
U^{(1)} &=& \bigg( \frac{\Phi_0}{2\pi}\bigg)^{\! \! 2} 
\bigg\lbrace \sum_{i=1,2}   \bigg[
\frac{\xi_i^{(0)} \xi_i^{(1)} }{L_{0i}}
+ \frac{(\xi_i^{(0)}-\varphi_i) \xi_i^{(1)} }{L_{{\rm j}i} }
 \nonumber \\
&-& \lambda \frac{ (\xi_i^{(0)}-\varphi_i)^4 }{24 \, L_{{\rm j}i} }
\bigg]  
+ \cos(\delta) \, \frac{ (\xi_1^{(0)} -\xi_2^{(0)})(\xi_1^{(1)}-\xi_2^{(1)})}{L_{\rm T}}
\nonumber \\
&+& \lambda^\prime \,  \sin(\delta) \,
\frac{(\xi_1^{(0)} -\xi_2^{(0)})^3}{6 L_{\rm T}} 
- \lambda^\prime \,  \cos(\delta) \,
\frac{(\xi_1^{(0)} -\xi_2^{(0)})^4}{24 L_{\rm T}}  \bigg\rbrace
\ \ \ \ \ \ \
\label{U1 definition}
\end{eqnarray}
is the anharmonic correction.

\begin{figure}
\includegraphics[width=9.0cm]{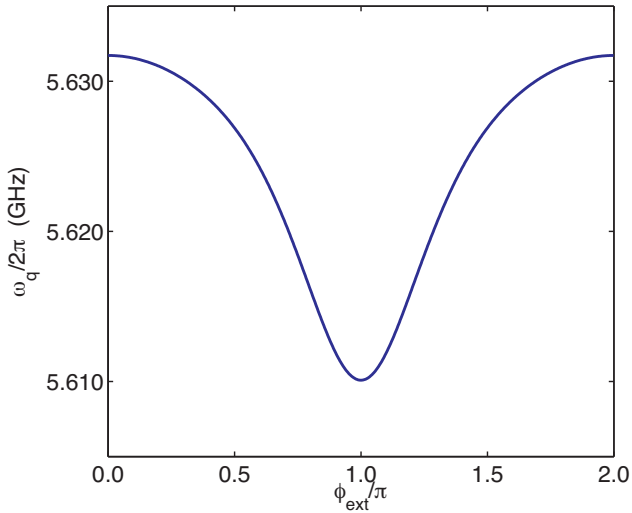} 
\caption{(Color online) Qubit frequency (\ref{qubit frequency}) as a function of external flux, assuming circuit parameters of Table~\ref{parameter table}. We see that $\omega_{\rm q}/2\pi$ varies by about $22 \, {\rm MHz}$ in this example.}
\label{qubit frequency figure}
\end{figure} 

\subsection{Coupling in the linearized model}

The Hamiltonian in the harmonic approximation is \begin{equation}
H = \sum_i H_i + \delta H,
\end{equation}
where [see (\ref{full hamiltonian expanded potential})]
\begin{equation}
H_i \equiv \bigg(\frac{2\pi}{\Phi_0}\bigg)^2 
\frac{p_i^2}{2C_i}  +  \bigg( \frac{\Phi_0}{2\pi}\bigg)^{\! \! 2} \frac{\varphi_i^2}{2 L_{{\rm q}i}} , \ \ \  (i=1,2)
\label{Hi}
\end{equation}
and
\begin{equation}
\delta H \equiv  \bigg( \frac{\Phi_0}{2\pi}\bigg)^{\! \! 2} \, \Gamma_{11} \, \varphi_1 \varphi_2.
\label{interaction Hamiltonian}
\end{equation}
The Hamiltonian (\ref{Hi}) describes a harmonic oscillator with flux-dependent frequency
\begin{equation}
\omega_{{\rm q}i} \equiv \sqrt{\frac{1}{L_{{\rm q}i} C_i}},
\label{qubit frequency}
\end{equation}
which is plotted in Fig.~\ref{qubit frequency figure} for the parameters of Table~\ref{parameter table}. Note that in the weak coupling analysis of Sec.~\ref{introduction section}, the qubit self-inductance (\ref{Lq in weak coupling}) and frequency are taken to be flux independent.

In Sec.~\ref{introduction section} we calculated the transverse coupling $g$ resulting from a $\varphi_1 \varphi_2$ interaction between a pair of identical classical harmonic oscillators [see (\ref{simple g})]. Here we derive the same result quantum mechanically (and for non-identical qubits). Let $|0\rangle_i$ and $|1\rangle_i$ be the ground and first excited state of $H_i$ (these are different than the eigenstates of the uncoupled qubits and they depend on $\phi_{\rm ext}$). Now we project the interaction term (\ref{interaction Hamiltonian}) into this basis. Each Josephson phase operator projects according to
\begin{eqnarray}
\varphi &\rightarrow& \begin{pmatrix} \varphi_{00} & \varphi_{01} \cr \varphi_{10} & \varphi_{11} 
\end{pmatrix} 
\nonumber \\ 
&=& \varphi_{01} \, \sigma^x - \big(\frac{ \varphi_{11}-\varphi_{00}}{2}\big) \, \sigma^z 
+ \big(\frac{ \varphi_{00}+\varphi_{11}}{2}\big) \, I,
\end{eqnarray}
where $\varphi_{mm'} \equiv \langle m | \varphi | m' \rangle.$
By symmetry $\varphi_{00} = \varphi_{11} = 0$, and because the potential in (\ref{Hi}) is parabolic,
\begin{equation}
\varphi_{01} =  \bigg(\frac{2 \pi}{\Phi_0}\bigg)
\sqrt{\frac{ \hbar L_{\rm q} \, \omega_{\rm q}}{2}}.
\label{dipole moment}
\end{equation}
Then we obtain, from (\ref{interaction Hamiltonian}), 
\begin{equation}
\delta H = g \, \sigma^x_1 \sigma^x_2,
\label{transverse interaction hamiltonian}
\end{equation} 
where
\begin{equation}
g = \frac{\hbar \Gamma_{11}\sqrt{ L_{{\rm q}1} L_{{\rm q}2}}}{2}  \, \sqrt{ \omega_{\rm q1} \omega_{\rm q2}},
\label{linear g}
\end{equation} 
which reduces to (\ref{simple g})
for symmetric qubits on resonance. The coupling strength (\ref{linear g}) is generally different than the simpler weak-coupling expression (\ref{simplified coupling}). However for the system parameters of Table~\ref{parameter table} they differ by no more than about $0.1 \, {\rm MHz}$.

\subsection{Nonlinear correction to transverse coupling}

To evaluate (\ref{U1 definition}) we will express (\ref{xi correction}) in terms of the coordinates $\varphi_1$ and $\varphi_2$. We note from (\ref{xi correction}) and (\ref{U1 definition}) that qubit nonlinearity $\lambda$ generates quartic terms in the corrections to the potential energy, whereas the coupler nonlinearity $\lambda^\prime$ generates both cubic and quartic terms. Although the complete expressions for $\xi_1^{(1)}$ and $\xi_2^{(1)}$ in terms of the $\varphi_i$ are quite complicated, they simplify when the circuit elements have identical parameters that satisfy 
\begin{equation}
L_0 \ll L_{\rm j} \ll L_{\rm T}.
\label{nonlinear correction assumptions}
\end{equation}
In this limit
\begin{eqnarray}
L_{\rm q} &\rightarrow& L_{\rm j}, \\
L_{\Sigma} &\rightarrow& L_0, \\
D &\rightarrow& \frac{1}{L_0^2}, \\
\alpha &\rightarrow& \frac{L_0}{L_{\rm j}}, \\
\beta &\rightarrow& \frac{\cos(\delta) \, L_0^2}{L_{\rm j} L_{\rm T}}, \\
\end{eqnarray}
and therefore
\begin{equation}
\beta \ll \alpha \ll 1.
\label{nonlinear correction limits}
\end{equation}
In this section we derive analytic expressions for the nonlinear corrections assuming (\ref{nonlinear correction assumptions}), which is a special case of the weak coupling assumption of Sec.~\ref{introduction section}. Using (\ref{nonlinear correction limits}) we find that
\begin{eqnarray}
\xi_1^{(1)} &\approx& \lambda \bigg( -\frac{\alpha}{6} \varphi_{1}^3 
+ \frac{\alpha\beta}{2} \varphi_{1}^2 \varphi_{2}
+  \frac{\beta^2}{2} \varphi_{1} \varphi_{2}^2
- \frac{\beta}{6} \varphi_{2}^3 \bigg)
\nonumber \\
&-& \lambda^\prime \
\frac{\alpha^2 L_0 \sin\delta}{2 L_{\rm T}} 
\bigg(   \varphi_{1}^2
-\varphi_{1} \varphi_{2} + \varphi_{2}^2 \bigg)
\label{xi1 expanded}
\end{eqnarray}
and
\begin{eqnarray}
\xi_2^{(1)} &\approx& \lambda \bigg( -\frac{\alpha}{6} \varphi_{2}^3 
+ \frac{\alpha\beta}{2} \varphi_{1} \varphi_{2}^2 
+  \frac{\beta^2}{2} \varphi_{1}^2 \varphi_{2} 
- \frac{\beta}{6} \varphi_{1}^3 \bigg)
\nonumber \\
&+& \lambda^\prime \
\frac{\alpha^2 L_0 \sin\delta}{2 L_{\rm T}} 
\bigg(   \varphi_{1}^2
- \varphi_{1} \varphi_{2} + \varphi_{2}^2 \bigg)
\label{xi2 expanded}.
\end{eqnarray}
These expressions are obtained by considering every term allowed by symmetry and approximating its coefficient by that of the dominant contribution (using $\lambda = \lambda^\prime = 1$). The correction (\ref{U1 definition}) is similarly obtained by assuming identical qubits and finding the largest contribution to every posssible term in the energy. The result is
\begin{eqnarray}
U^{(1)} &=& \bigg( \frac{\Phi_0}{2\pi}\bigg)^{\! \! 2} 
\bigg[ 
\lambda \, \Gamma_{04} \, \big( \varphi_1^4 + \varphi_2^4 \big)
+ \lambda^\prime \, \Gamma_{03} \, \big( \varphi_1^3 - \varphi_2^3 \big) 
\nonumber \\
&+& \lambda \, \Gamma_{13} \, \big( \varphi_1 \varphi_2^3 + \varphi_1^3 \varphi_2 \big)
+ \lambda^\prime \, \Gamma_{12} \, \big( \varphi_1 \varphi_2^2 - \varphi_1^2 \varphi_2 \big)
\nonumber \\
&+& \lambda \, \Gamma_{22} \, \varphi_1^2 \varphi_2^2 
\bigg],
\label{U1}
\end{eqnarray}
where
\begin{eqnarray}
\Gamma_{04} &=& -\frac{1}{24 L_{\rm j}}  , \\
\Gamma_{03} &=& \frac{\alpha^3\sin\delta}{6 L_{\rm T}}, 
 \\
\Gamma_{13} &=& \frac{\alpha^2\cos\delta}{6 L_{\rm T}}    ,\\
\Gamma_{12} &=&  \frac{\alpha^3\sin\delta}{2 L_{\rm T}} ,
   \\
\Gamma_{22} &=& \alpha \beta \bigg( \frac{\beta}{L_0} - \frac{\alpha \cos\delta}{L_{\rm T}}\bigg) .
\end{eqnarray}

\begin{figure}
\includegraphics[width=9.0cm]{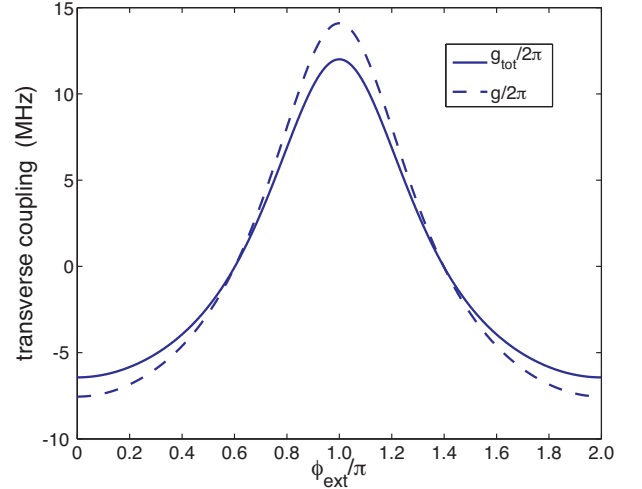} 
\caption{(Color online) Coupling strength in the perturbative nonlinear approximation for system parameters given in Table~\ref{parameter table}. The dashed line is coupling (\ref{linear g}) in the linearized model.}
\label{nonlinear coupling figure}
\end{figure} 

The dominant nonlinear correction to the transverse coupling is 
\begin{equation}
\delta g = \bigg( \frac{\Phi_0}{2\pi}\bigg)^{\! \! 2} 
\Gamma_{13} \,
\langle 01 | \varphi_1 \varphi_2^3 + \varphi_1^3 \varphi_2 | 10\rangle.
\label{delta g definition}
\end{equation}
To evaluate ({\ref{delta g definition}) note that
$ \langle 01 | \varphi_1 \varphi_2^3 + \varphi_1^3 \varphi_2 | 10\rangle
= 2  \varphi_{01} \, \langle 0 | \varphi^3 | 1 \rangle, $
where $\varphi_{01} $ is defined in (\ref{dipole moment}) and
\begin{equation}
 \langle 0 | \varphi^3 | 1 \rangle = 3 \,
\bigg(\frac{2 \pi}{\Phi_0}\bigg)^{\! 3}
\bigg(\frac{\hbar L_{\rm q} \, \omega_{\rm q}}{2}\bigg)^{\! \frac{3}{2}} \! .
\end{equation}
Then ({\ref{delta g definition}) can be written as
\begin{eqnarray}
\delta g &=& \frac{3}{2} \, \Gamma_{13} \, 
\bigg( \frac{\hbar \omega_{\rm q} L_{\rm q}}{\Phi_0/2 
\pi} \bigg)^{\! 2} \\
&=&
\cos(\delta)
\frac{\pi^2\alpha^2 L_{\rm j}}{2L_{\rm T}}
\bigg( \frac{\hbar \omega_{\rm q}}{\Phi_0^2/2 
L_{\rm j}} \bigg) \, \hbar \omega_{\rm q} .
\label{delta g expression}
\end{eqnarray}

The total transverse coupling 
\begin{equation}
g_{\rm tot} \equiv g + \delta g
\label{total transverse coupling definition}
\end{equation}
obtained from (\ref{linear g}) and (\ref{delta g expression}) is plotted in Fig.~\ref{nonlinear coupling figure}. Note that nonlinear contribution zeros precisely where the linear coupling does, and that the correction always suppresses the magnitude of the coupling. The amount of coupling suppression can be simply quantified by writing (\ref{total transverse coupling definition}) as
\begin{equation}
g_{\rm tot}  = \zeta \, g, 
\ \ {\rm where} \ \ \zeta \equiv 1 + \frac{\delta g}{g}.
\label{zeta definition}
\end{equation}
We emphasize that $g$ in (\ref{zeta definition}) refers to the coupling (\ref{simplified coupling}) or (\ref{linear g}) for the linearized circuit. To estimate $\zeta$ we again 
assume (\ref{nonlinear correction assumptions}), which leads to
\begin{equation}
\zeta \approx 1 - \pi^2 \bigg( \frac{\hbar \omega_{\rm q}}{\Phi_0^2/2 
L_{\rm j}} \bigg) = 0.852,
\end{equation}
using a qubit frequency of $5.62 \, {\rm GHz}$ and the value of $L_{\rm j}$ from Table~\ref{parameter table}.
Therefore we find that qubit nonlinearity suppresses the transverse coupling by about $15\%,$ and that the effects of coupler nonlinearity (corrections proportional to $\lambda^\prime$) are negligible in the parameter regime considered.

To validate the perturbative correction (\ref{delta g expression}) we compare, in Fig.~\ref{g exact diagonalization figure}, the splitting $2 | g_{\rm tot} |$ between the symmetric and antisymmetric eigenstates to the fully nonlinear result obtained by exact diagonalization. We find that the analytic approximation developed here is in very good agreement with the numerical results. It can be shown that the small differences arise not from the replacement of the cosine potentials by their quadratic plus quartic expansions, but from (i) keeping only the terms first order in $\lambda$ and $\lambda^\prime$ in the subsequent analysis, and (ii) assuming the limit (\ref{nonlinear correction assumptions}).

\section{Diagonal coupling}
\label{diagonal coupling section}

The coupler circuit of Fig.~\ref{coupler circuit figure} also produces a small diagonal qubit-qubit interaction of the form
\begin{equation}
\delta H = J \sigma^z_1 \sigma^z_2.
\label{diagonal interaction hamiltonian}
\end{equation}
In this section we calculate $J$, analytically and numerically, by relating it to the exact eigenstates of the coupled qubit system \cite{PintoPRB10}, 
\begin{equation}
J = \frac{E_{11} - (E_{+} + E_{-})+ E_{00}}{4},
\label{J definition}
\end{equation}
and throughout this section we assume resonantly tuned qubits. Here $E_{11}$ is the energy of the $|11\rangle$ state,
\begin{equation} 
E_{\pm} = \omega_{\rm q} \pm |g| + E_{00}
\end{equation}
are the energies of the single-excitation eigenstates, with $\omega_{\rm q}$ the frequency of the uncoupled qubits, and $E_{00}$ is the ground state energy. Note that $J$ is to be computed in the presence of the {\it total} transverse interaction
\begin{equation}
\delta H = g \, \sigma^x_1 \sigma^x_2,
\label{total transverse interaction}
\end{equation}
where in this section we write $g_{\rm tot}$ [defined in (\ref{total transverse coupling definition})] simply as $g$.

\begin{figure}
\includegraphics[width=9.0cm]{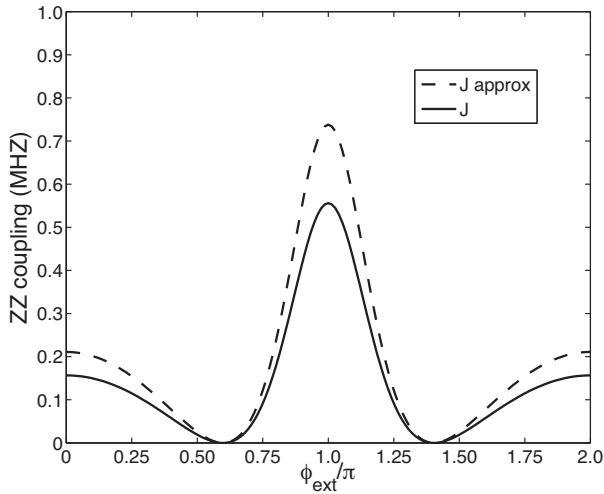} 
\caption{Diagonal coupling strength (\ref{J definition}) computed by exact diagonalization, and the approximation (\ref{J approximation}).}
\label{J exact diagonalization figure}
\end{figure} 

Two types of effects contribute to the total diagonal coupling $J$. The dominant mechanism comes from states outside of the qubit subspace and is caused by the repulsion of $|11\rangle$ by the $|02\rangle$ and $|20\rangle$ eigenstates. These states differ in energy by the qubit anharmonicity
\begin{equation}
\eta \equiv (E_1 - E_0) - (E_2 - E_1).
\label{eta definition}
\end{equation}
Referring to the nonlinear Hamiltonian (\ref{U1}), this contribution to $J$ results from the terms proportional to $\Gamma_{04}$ and $\Gamma_{03}$, which generate qubit anharmonicity, in the presence of a transverse interaction.

\begin{figure}
\includegraphics[width=9.0cm]{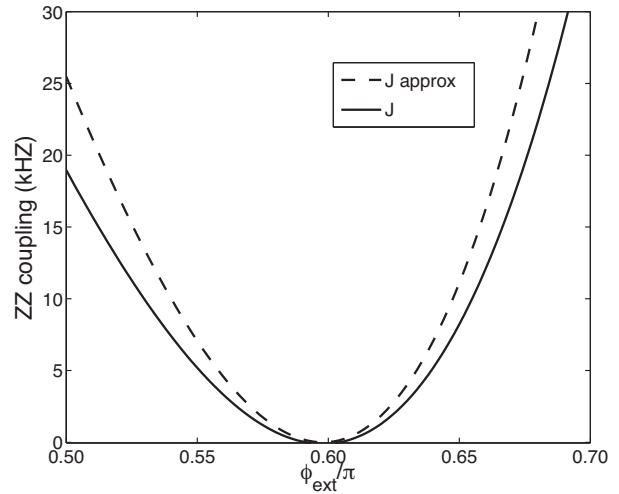} 
\caption{Expanded view of Fig.~\ref{J exact diagonalization figure} (note kHz frequency scale).}
\label{J exact diagonalization expanded figure}
\end{figure} 

We can estimate this effect by considering the second-order correction to the energy of the $|11\rangle$ state resulting from the transverse interaction, which is
\begin{equation}
\delta E_{11} \approx 2 \times \! \frac{(\sqrt{2} g)^2}{\eta},
\label{E11 shift}
\end{equation}
assuming harmonic oscillator eigenfunctions. The factor of 2 in (\ref{E11 shift}) comes from the contributions by both $|02\rangle$ and $|20\rangle$. Then the $\sigma^z \otimes \sigma^z$ coupling strength is simply
\begin{equation}
J \approx \frac{g^2}{\eta}.
\label{J approximation}
\end{equation}

A few remarks about (\ref{J approximation}) are in order: The diagonal coupling resulting from the $|2\rangle$ state repulsion effect is always positive, and it zeros when the transverse coupling does. However other contributions to $J$ (see below) can have either sign. Also, the use of harmonic oscillator eigenfunctions will slightly overestimate the $E_{11}$ repulsion and hence $J$. Finally, the anharmonicity and size of $\eta$ generated by the terms proportional to $\Gamma_{04}$ (which are dominant and flux independent) and $\Gamma_{03}$ (which depends on $\Phi_{\rm ext}$) is an approximation, so in (\ref{J approximation}) we instead prefer to use an exactly calculated (or measured) value, which is approximately $213 \, {\rm MHz}$ for uncoupled qubits with parameters of Table~\ref{parameter table}. 

The $\sigma^z \otimes \sigma^z$ coupling strength (\ref{J definition}) for a system with parameters of Table~\ref{parameter table} is shown in Fig.~\ref{J exact diagonalization figure}, along with the approximation (\ref{J approximation}). Here (\ref{J definition}) is computed by exact diagonalization and is shown in the solid curve. The approximation (\ref{J approximation}) is evaluated by using the exact diagonalization result for the total transverse coupling $g$, with $\eta/2\pi \! = \! 213 \, {\rm MHz}$, and is shown in the dashed curve.

Although the approximation (\ref{J approximation}) necessarily zeros when $g$ does, the exact value calculated from (\ref{J definition}) does not have to. In Fig.~\ref{J exact diagonalization expanded figure} we show an expanded view of Fig.~\ref{J exact diagonalization figure} near a minimum. We find that the $\sigma^z \otimes \sigma^z$ coupling strength (\ref{J definition}) calculated by exact diagonalization does reach a negative value of $-110 \, {\rm Hz}$, but this tiny value may not be reliable given our numerical accuracy.

The second type of effects contributing to $J$ result from the interaction terms proportional to $\Gamma_{13}$, $\Gamma_{12}$, and $\Gamma_{22}$ in (\ref{U1}). The $\Gamma_{22}$ terms make the largest contribution to $J$, because they are the only ones that survive when the small anharmonic corrections to the qubit {\it eigenfunctions} are neglected. To estimate the $\Gamma_{22}$ contributions we project the $\varphi_i^2$ operators as
\begin{equation}
\varphi^2 \rightarrow \begin{pmatrix} 
\langle 0 | \varphi^2 | 0 \rangle  & \langle 0 | \varphi^2 | 1 \rangle \cr \langle 1 | \varphi^2 | 0 \rangle & \langle 1 | \varphi^2 | 1 \rangle 
\end{pmatrix} 
\approx 
\bigg( \frac{2 \pi}{\Phi_0} \bigg)
\hbar \omega_{\rm q} L_{\rm q} \times \big( I - \textstyle{\frac{1}{2}} \sigma^z \big),
\label{phi squared projection}
\end{equation}
where $I$ is the identity matrix and in the second step we have assumed harmonic eigenfunctions. This leads to an additional contribution \begin{equation}
J =  \Gamma_{22} \, \bigg( \frac{2 \pi}{\Phi_0} \bigg)^2 \bigg( \frac{\hbar \omega_{\rm q} L_{\rm q}}{2} \bigg)^2 \! \! ,
\label{J subdominant}
\end{equation}
which is always much smaller than (\ref{J approximation}) and also zeros when $g$ does. The subdominant contribution (\ref{J subdominant}) is plotted in Fig.~\ref{J subdominant figure} using the parameters of Table~\ref{parameter table}. 

\begin{figure}
\includegraphics[width=9.0cm]{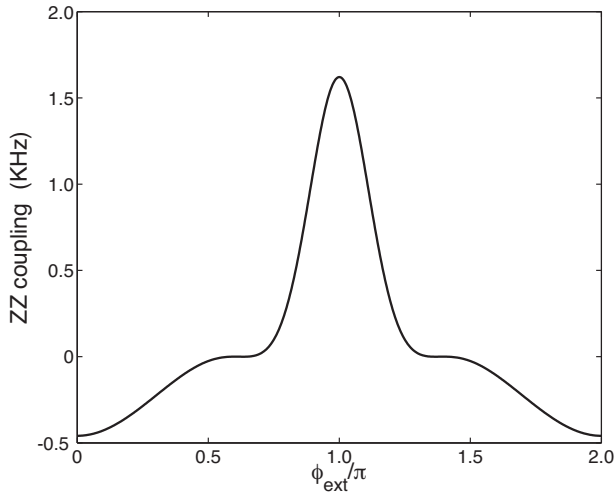} 
\caption{Subdominant contribution (\ref{J subdominant}) to $J$, versus flux.}
\label{J subdominant figure}
\end{figure} 

\begin{acknowledgments}

This research was funded by the US Office of the Director of National Intelligence (ODNI), Intelligence Advanced Research Projects Activity (IARPA), through the US Army Research Office grant No.~W911NF-10-1-0334. All statements of fact, opinion or conclusions contained herein are those of the authors and should not be construed as representing the official views or policies of IARPA, the ODNI, or the US Government. We thank Alexander Korotkov for useful discussions.

\end{acknowledgments}

\bibliography{/Users/mgeller/bibliographies/algorithms,/Users/mgeller/bibliographies/dwave,/Users/mgeller/bibliographies/control,/Users/mgeller/bibliographies/error_correction,/Users/mgeller/bibliographies/general,/Users/mgeller/bibliographies/group,/Users/mgeller/bibliographies/ions,/Users/mgeller/bibliographies/math,/Users/mgeller/bibliographies/nmr,/Users/mgeller/bibliographies/optics,/Users/mgeller/bibliographies/simulation,/Users/mgeller/bibliographies/superconductors,/Users/mgeller/bibliographies/surface_code,endnotes}

\begin{thebibliography}{24}%
\makeatletter
\providecommand \@ifxundefined [1]{%
 \@ifx{#1\undefined}
}%
\providecommand \@ifnum [1]{%
 \ifnum #1\expandafter \@firstoftwo
 \else \expandafter \@secondoftwo
 \fi
}%
\providecommand \@ifx [1]{%
 \ifx #1\expandafter \@firstoftwo
 \else \expandafter \@secondoftwo
 \fi
}%
\providecommand \natexlab [1]{#1}%
\providecommand \enquote  [1]{``#1''}%
\providecommand \bibnamefont  [1]{#1}%
\providecommand \bibfnamefont [1]{#1}%
\providecommand \citenamefont [1]{#1}%
\providecommand \href@noop [0]{\@secondoftwo}%
\providecommand \href [0]{\begingroup \@sanitize@url \@href}%
\providecommand \@href[1]{\@@startlink{#1}\@@href}%
\providecommand \@@href[1]{\endgroup#1\@@endlink}%
\providecommand \@sanitize@url [0]{\catcode `\\12\catcode `\$12\catcode
  `\&12\catcode `\#12\catcode `\^12\catcode `\_12\catcode `\%12\relax}%
\providecommand \@@startlink[1]{}%
\providecommand \@@endlink[0]{}%
\providecommand \url  [0]{\begingroup\@sanitize@url \@url }%
\providecommand \@url [1]{\endgroup\@href {#1}{\urlprefix }}%
\providecommand \urlprefix  [0]{URL }%
\providecommand \Eprint [0]{\href }%
\@ifxundefined \urlstyle {%
  \providecommand \doi  [0]{\begingroup \@sanitize@url \@doi}%
  \providecommand \@doi [1]{\endgroup \@@startlink {\doibase
  #1}doi:\discretionary {}{}{}#1\@@endlink }%
}{%
  \providecommand \doi  [0]{doi:\discretionary{}{}{}\begingroup
  \urlstyle{rm}\Url }%
}%
\providecommand \doibase [0]{http://dx.doi.org/}%
\providecommand \Doi [0]{\begingroup \@sanitize@url \@Doi }%
\providecommand \@Doi  [1]{\endgroup\@@startlink{\doibase#1}\@@Doi}%
\providecommand \@@Doi [1]{#1\@@endlink}%
\providecommand \selectlanguage [0]{\@gobble}%
\providecommand \bibinfo  [0]{\@secondoftwo}%
\providecommand \bibfield  [0]{\@secondoftwo}%
\providecommand \translation [1]{[#1]}%
\providecommand \BibitemOpen [0]{}%
\providecommand \bibitemStop [0]{}%
\providecommand \bibitemNoStop [0]{.\EOS\space}%
\providecommand \EOS [0]{\spacefactor3000\relax}%
\providecommand \BibitemShut  [1]{\csname bibitem#1\endcsname}%
\bibitem [{\citenamefont {Barends}\ \emph {et~al.}(2013)\citenamefont
  {Barends}, \citenamefont {Kelly}, \citenamefont {Megrant}, \citenamefont
  {Sank}, \citenamefont {Jeffrey}, \citenamefont {Chen}, \citenamefont {Yin},
  \citenamefont {Chiaro}, \citenamefont {Mutus}, \citenamefont {Neill},
  \citenamefont {O'Malley}, \citenamefont {Roushan}, \citenamefont {Wenner},
  \citenamefont {White}, \citenamefont {Cleland},\ and\ \citenamefont
  {Martinis}}]{BarendsPRL13}%
  \BibitemOpen
  \bibfield  {author} {\bibinfo {author} {\bibfnamefont {R.}~\bibnamefont
  {Barends}}, \bibinfo {author} {\bibfnamefont {J.}~\bibnamefont {Kelly}},
  \bibinfo {author} {\bibfnamefont {A.}~\bibnamefont {Megrant}}, \bibinfo
  {author} {\bibfnamefont {D.}~\bibnamefont {Sank}}, \bibinfo {author}
  {\bibfnamefont {E.}~\bibnamefont {Jeffrey}}, \bibinfo {author} {\bibfnamefont
  {Y.}~\bibnamefont {Chen}}, \bibinfo {author} {\bibfnamefont {Y.}~\bibnamefont
  {Yin}}, \bibinfo {author} {\bibfnamefont {B.}~\bibnamefont {Chiaro}},
  \bibinfo {author} {\bibfnamefont {J.}~\bibnamefont {Mutus}}, \bibinfo
  {author} {\bibfnamefont {C.}~\bibnamefont {Neill}}, \bibinfo {author}
  {\bibfnamefont {P.}~\bibnamefont {O'Malley}}, \bibinfo {author}
  {\bibfnamefont {P.}~\bibnamefont {Roushan}}, \bibinfo {author} {\bibfnamefont
  {J.}~\bibnamefont {Wenner}}, \bibinfo {author} {\bibfnamefont {T.~C.}\
  \bibnamefont {White}}, \bibinfo {author} {\bibfnamefont {A.~N.}\ \bibnamefont
  {Cleland}}, \ and\ \bibinfo {author} {\bibfnamefont {J.~M.}\ \bibnamefont
  {Martinis}},\ }\href@noop {} {\bibfield  {journal} {\bibinfo  {journal}
  {Phys. Rev. Lett.},\ }\textbf {\bibinfo {volume} {111}},\ \bibinfo {pages}
  {080502} (\bibinfo {year} {2013})}\BibitemShut {NoStop}%
\bibitem [{\citenamefont {Barends}\ \emph {et~al.}(2014)\citenamefont
  {Barends}, \citenamefont {Kelly}, \citenamefont {Megrant}, \citenamefont
  {Veitia}, \citenamefont {Sank}, \citenamefont {Jeffrey}, \citenamefont
  {White}, \citenamefont {Mutus}, \citenamefont {Fowler}, \citenamefont
  {Campbell}, \citenamefont {Chen}, \citenamefont {Chen}, \citenamefont
  {Chiaro}, \citenamefont {Dunsworth}, \citenamefont {Neill}, \citenamefont
  {O$^\prime$~Malley}, \citenamefont {Roushan}, \citenamefont {Vainsencher},
  \citenamefont {Wenner}, \citenamefont {Korotkov}, \citenamefont {Cleland},\
  and\ \citenamefont {Martinis}}]{BarendsNat14}%
  \BibitemOpen
  \bibfield  {author} {\bibinfo {author} {\bibfnamefont {R.}~\bibnamefont
  {Barends}}, \bibinfo {author} {\bibfnamefont {J.}~\bibnamefont {Kelly}},
  \bibinfo {author} {\bibfnamefont {A.}~\bibnamefont {Megrant}}, \bibinfo
  {author} {\bibfnamefont {A.}~\bibnamefont {Veitia}}, \bibinfo {author}
  {\bibfnamefont {D.}~\bibnamefont {Sank}}, \bibinfo {author} {\bibfnamefont
  {E.}~\bibnamefont {Jeffrey}}, \bibinfo {author} {\bibfnamefont {T.~C.}\
  \bibnamefont {White}}, \bibinfo {author} {\bibfnamefont {J.}~\bibnamefont
  {Mutus}}, \bibinfo {author} {\bibfnamefont {A.~G.}\ \bibnamefont {Fowler}},
  \bibinfo {author} {\bibfnamefont {B.}~\bibnamefont {Campbell}}, \bibinfo
  {author} {\bibfnamefont {Y.}~\bibnamefont {Chen}}, \bibinfo {author}
  {\bibfnamefont {Z.}~\bibnamefont {Chen}}, \bibinfo {author} {\bibfnamefont
  {B.}~\bibnamefont {Chiaro}}, \bibinfo {author} {\bibfnamefont
  {A.}~\bibnamefont {Dunsworth}}, \bibinfo {author} {\bibfnamefont
  {C.}~\bibnamefont {Neill}}, \bibinfo {author} {\bibfnamefont
  {P.}~\bibnamefont {O$^\prime$~Malley}}, \bibinfo {author} {\bibfnamefont
  {P.}~\bibnamefont {Roushan}}, \bibinfo {author} {\bibfnamefont
  {A.}~\bibnamefont {Vainsencher}}, \bibinfo {author} {\bibfnamefont
  {J.}~\bibnamefont {Wenner}}, \bibinfo {author} {\bibfnamefont {A.~N.}\
  \bibnamefont {Korotkov}}, \bibinfo {author} {\bibfnamefont {A.~N.}\
  \bibnamefont {Cleland}}, \ and\ \bibinfo {author} {\bibfnamefont {J.~M.}\
  \bibnamefont {Martinis}},\ }\href@noop {} {\bibfield  {journal} {\bibinfo
  {journal} {Nature (London)},\ }\textbf {\bibinfo {volume} {508}},\ \bibinfo
  {pages} {500} (\bibinfo {year} {2014})}\BibitemShut {NoStop}%
\bibitem [{\citenamefont {Geller}\ \emph {et~al.}()\citenamefont {Geller},
  \citenamefont {Martinis}, \citenamefont {Sornborger}, \citenamefont
  {Stancil}, \citenamefont {Pritchett},\ and\ \citenamefont
  {Galiautdinov}}]{GellerMartinisSornborgerEtalPre12}%
  \BibitemOpen
  \bibfield  {author} {\bibinfo {author} {\bibfnamefont {M.~R.}\ \bibnamefont
  {Geller}}, \bibinfo {author} {\bibfnamefont {J.~M.}\ \bibnamefont
  {Martinis}}, \bibinfo {author} {\bibfnamefont {A.~T.}\ \bibnamefont
  {Sornborger}}, \bibinfo {author} {\bibfnamefont {P.~C.}\ \bibnamefont
  {Stancil}}, \bibinfo {author} {\bibfnamefont {E.~J.}\ \bibnamefont
  {Pritchett}}, \ and\ \bibinfo {author} {\bibfnamefont {A.}~\bibnamefont
  {Galiautdinov}},\ }\href@noop {} {\enquote {\bibinfo {title} {Universal
  quantum simulation with pre-threshold superconducting qubits:
  {S}ingle-excitation subspace method},}\ }\bibinfo {note}
  {{a}rXiv:1210.5260}\BibitemShut {NoStop}%
\bibitem [{\citenamefont {Makhlin}\ \emph {et~al.}(1999)\citenamefont
  {Makhlin}, \citenamefont {Sch\"on},\ and\ \citenamefont
  {Shnirman}}]{MakhlinNat99}%
  \BibitemOpen
  \bibfield  {author} {\bibinfo {author} {\bibfnamefont {Y.}~\bibnamefont
  {Makhlin}}, \bibinfo {author} {\bibfnamefont {G.}~\bibnamefont {Sch\"on}}, \
  and\ \bibinfo {author} {\bibfnamefont {A.}~\bibnamefont {Shnirman}},\
  }\href@noop {} {\bibfield  {journal} {\bibinfo  {journal} {Nature (London)},\
  }\textbf {\bibinfo {volume} {398}},\ \bibinfo {pages} {305} (\bibinfo {year}
  {1999})}\BibitemShut {NoStop}%
\bibitem [{\citenamefont {Blais}\ \emph {et~al.}(2003)\citenamefont {Blais},
  \citenamefont {Massen van~den Brink},\ and\ \citenamefont
  {Zagoskin}}]{BlaisPRL03}%
  \BibitemOpen
  \bibfield  {author} {\bibinfo {author} {\bibfnamefont {A.}~\bibnamefont
  {Blais}}, \bibinfo {author} {\bibfnamefont {A.}~\bibnamefont {Massen van~den
  Brink}}, \ and\ \bibinfo {author} {\bibfnamefont {A.~M.}\ \bibnamefont
  {Zagoskin}},\ }\href@noop {} {\bibfield  {journal} {\bibinfo  {journal}
  {Phys. Rev. Lett.},\ }\textbf {\bibinfo {volume} {90}},\ \bibinfo {pages}
  {127901} (\bibinfo {year} {2003})}\BibitemShut {NoStop}%
\bibitem [{\citenamefont {Plourde}\ \emph {et~al.}(2004)\citenamefont
  {Plourde}, \citenamefont {Zhang}, \citenamefont {Whaley}, \citenamefont
  {Wilhelm}, \citenamefont {Robertson}, \citenamefont {Hime}, \citenamefont
  {Linzen}, \citenamefont {Reichardt}, \citenamefont {Wu},\ and\ \citenamefont
  {Clarke}}]{PlourdePRB04}%
  \BibitemOpen
  \bibfield  {author} {\bibinfo {author} {\bibfnamefont {B.~L.~T.}\
  \bibnamefont {Plourde}}, \bibinfo {author} {\bibfnamefont {J.}~\bibnamefont
  {Zhang}}, \bibinfo {author} {\bibfnamefont {K.~B.}\ \bibnamefont {Whaley}},
  \bibinfo {author} {\bibfnamefont {F.~K.}\ \bibnamefont {Wilhelm}}, \bibinfo
  {author} {\bibfnamefont {T.~L.}\ \bibnamefont {Robertson}}, \bibinfo {author}
  {\bibfnamefont {T.}~\bibnamefont {Hime}}, \bibinfo {author} {\bibfnamefont
  {S.}~\bibnamefont {Linzen}}, \bibinfo {author} {\bibfnamefont {P.~A.}\
  \bibnamefont {Reichardt}}, \bibinfo {author} {\bibfnamefont {C.-E.}\
  \bibnamefont {Wu}}, \ and\ \bibinfo {author} {\bibfnamefont {J.}~\bibnamefont
  {Clarke}},\ }\href@noop {} {\bibfield  {journal} {\bibinfo  {journal} {Phys.
  Rev. B},\ }\textbf {\bibinfo {volume} {70}} (\bibinfo {year}
  {2004})}\BibitemShut {NoStop}%
\bibitem [{\citenamefont {Lantz}\ \emph {et~al.}(2004)\citenamefont {Lantz},
  \citenamefont {Wallquist}, \citenamefont {Shumeiko},\ and\ \citenamefont
  {Wendin}}]{LantzPRB04}%
  \BibitemOpen
  \bibfield  {author} {\bibinfo {author} {\bibfnamefont {J.}~\bibnamefont
  {Lantz}}, \bibinfo {author} {\bibfnamefont {M.}~\bibnamefont {Wallquist}},
  \bibinfo {author} {\bibfnamefont {V.~S.}\ \bibnamefont {Shumeiko}}, \ and\
  \bibinfo {author} {\bibfnamefont {G.}~\bibnamefont {Wendin}},\ }\href@noop {}
  {\bibfield  {journal} {\bibinfo  {journal} {Phys. Rev. B},\ }\textbf
  {\bibinfo {volume} {70}},\ \bibinfo {pages} {140507} (\bibinfo {year}
  {2004})}\BibitemShut {NoStop}%
\bibitem [{\citenamefont {Grajcar}\ \emph {et~al.}(2006)\citenamefont
  {Grajcar}, \citenamefont {Liu}, \citenamefont {Nori},\ and\ \citenamefont
  {Zagoskin}}]{GrajcarPRB06}%
  \BibitemOpen
  \bibfield  {author} {\bibinfo {author} {\bibfnamefont {M.}~\bibnamefont
  {Grajcar}}, \bibinfo {author} {\bibfnamefont {Y.-X.}\ \bibnamefont {Liu}},
  \bibinfo {author} {\bibfnamefont {F.}~\bibnamefont {Nori}}, \ and\ \bibinfo
  {author} {\bibfnamefont {A.~M.}\ \bibnamefont {Zagoskin}},\ }\href@noop {}
  {\bibfield  {journal} {\bibinfo  {journal} {Phys. Rev. B},\ }\textbf
  {\bibinfo {volume} {74}},\ \bibinfo {pages} {172505} (\bibinfo {year}
  {2006})}\BibitemShut {NoStop}%
\bibitem [{\citenamefont {Hime}\ \emph {et~al.}(2006)\citenamefont {Hime},
  \citenamefont {Reichardt}, \citenamefont {Plourde}, \citenamefont
  {Robertson}, \citenamefont {Wu}, \citenamefont {Ustinov},\ and\ \citenamefont
  {Clarke}}]{HimeSci06}%
  \BibitemOpen
  \bibfield  {author} {\bibinfo {author} {\bibfnamefont {T.}~\bibnamefont
  {Hime}}, \bibinfo {author} {\bibfnamefont {P.~A.}\ \bibnamefont {Reichardt}},
  \bibinfo {author} {\bibfnamefont {B.~L.~T.}\ \bibnamefont {Plourde}},
  \bibinfo {author} {\bibfnamefont {T.~L.}\ \bibnamefont {Robertson}}, \bibinfo
  {author} {\bibfnamefont {C.-E.}\ \bibnamefont {Wu}}, \bibinfo {author}
  {\bibfnamefont {V.~A.}\ \bibnamefont {Ustinov}}, \ and\ \bibinfo {author}
  {\bibfnamefont {J.}~\bibnamefont {Clarke}},\ }\href@noop {} {\bibfield
  {journal} {\bibinfo  {journal} {Science},\ }\textbf {\bibinfo {volume}
  {314}},\ \bibinfo {pages} {1427} (\bibinfo {year} {2006})}\BibitemShut
  {NoStop}%
\bibitem [{\citenamefont {van~der Ploeg}\ \emph {et~al.}(2007)\citenamefont
  {van~der Ploeg}, \citenamefont {Izmalkov}, \citenamefont {Maassen van~den
  Brink}, \citenamefont {Hubner}, \citenamefont {Grajcar}, \citenamefont
  {Il'ichev}, \citenamefont {Meyer},\ and\ \citenamefont
  {Zagoskin}}]{vanderPloegPRL07}%
  \BibitemOpen
  \bibfield  {author} {\bibinfo {author} {\bibfnamefont {S.~H.~W.}\
  \bibnamefont {van~der Ploeg}}, \bibinfo {author} {\bibfnamefont
  {A.}~\bibnamefont {Izmalkov}}, \bibinfo {author} {\bibfnamefont
  {A.}~\bibnamefont {Maassen van~den Brink}}, \bibinfo {author} {\bibfnamefont
  {U.}~\bibnamefont {Hubner}}, \bibinfo {author} {\bibfnamefont
  {M.}~\bibnamefont {Grajcar}}, \bibinfo {author} {\bibfnamefont
  {E.}~\bibnamefont {Il'ichev}}, \bibinfo {author} {\bibfnamefont {H.-G.}\
  \bibnamefont {Meyer}}, \ and\ \bibinfo {author} {\bibfnamefont {A.~M.}\
  \bibnamefont {Zagoskin}},\ }\href@noop {} {\bibfield  {journal} {\bibinfo
  {journal} {Phys. Rev. Lett.},\ }\textbf {\bibinfo {volume} {98}},\ \bibinfo
  {pages} {057004} (\bibinfo {year} {2007})}\BibitemShut {NoStop}%
\bibitem [{\citenamefont {Niskanen}\ \emph {et~al.}(2007)\citenamefont
  {Niskanen}, \citenamefont {Harrabi}, \citenamefont {Yoshihara}, \citenamefont
  {Nakamura}, \citenamefont {Lloyd},\ and\ \citenamefont
  {Tsai}}]{NiskanenSci07}%
  \BibitemOpen
  \bibfield  {author} {\bibinfo {author} {\bibfnamefont {A.~O.}\ \bibnamefont
  {Niskanen}}, \bibinfo {author} {\bibfnamefont {K.}~\bibnamefont {Harrabi}},
  \bibinfo {author} {\bibfnamefont {F.}~\bibnamefont {Yoshihara}}, \bibinfo
  {author} {\bibfnamefont {Y.}~\bibnamefont {Nakamura}}, \bibinfo {author}
  {\bibfnamefont {S.}~\bibnamefont {Lloyd}}, \ and\ \bibinfo {author}
  {\bibfnamefont {J.~S.}\ \bibnamefont {Tsai}},\ }\href@noop {} {\bibfield
  {journal} {\bibinfo  {journal} {Science},\ }\textbf {\bibinfo {volume}
  {316}},\ \bibinfo {pages} {723} (\bibinfo {year} {2007})}\BibitemShut
  {NoStop}%
\bibitem [{\citenamefont {Harris}\ \emph {et~al.}(2007)\citenamefont {Harris},
  \citenamefont {Berkley}, \citenamefont {Johnson}, \citenamefont {Bunyk},
  \citenamefont {Govorkov}, \citenamefont {Thom}, \citenamefont {Uchaikin},
  \citenamefont {Wilson}, \citenamefont {Chung}, \citenamefont {Holtham},
  \citenamefont {Biamonte}, \citenamefont {Smirnov}, \citenamefont {Amin},\
  and\ \citenamefont {Maassen van~den Brink}}]{HarrisPRL07}%
  \BibitemOpen
  \bibfield  {author} {\bibinfo {author} {\bibfnamefont {R.}~\bibnamefont
  {Harris}}, \bibinfo {author} {\bibfnamefont {A.~J.}\ \bibnamefont {Berkley}},
  \bibinfo {author} {\bibfnamefont {M.~W.}\ \bibnamefont {Johnson}}, \bibinfo
  {author} {\bibfnamefont {P.}~\bibnamefont {Bunyk}}, \bibinfo {author}
  {\bibfnamefont {S.}~\bibnamefont {Govorkov}}, \bibinfo {author}
  {\bibfnamefont {M.~C.}\ \bibnamefont {Thom}}, \bibinfo {author}
  {\bibfnamefont {S.}~\bibnamefont {Uchaikin}}, \bibinfo {author}
  {\bibfnamefont {A.~B.}\ \bibnamefont {Wilson}}, \bibinfo {author}
  {\bibfnamefont {J.}~\bibnamefont {Chung}}, \bibinfo {author} {\bibfnamefont
  {E.}~\bibnamefont {Holtham}}, \bibinfo {author} {\bibfnamefont {J.~D.}\
  \bibnamefont {Biamonte}}, \bibinfo {author} {\bibfnamefont {A.~Y.}\
  \bibnamefont {Smirnov}}, \bibinfo {author} {\bibfnamefont {M.~H.~S.}\
  \bibnamefont {Amin}}, \ and\ \bibinfo {author} {\bibfnamefont
  {A.}~\bibnamefont {Maassen van~den Brink}},\ }\href@noop {} {\bibfield
  {journal} {\bibinfo  {journal} {Phys. Rev. Lett.},\ }\textbf {\bibinfo
  {volume} {98}},\ \bibinfo {pages} {177001} (\bibinfo {year}
  {2007})}\BibitemShut {NoStop}%
\bibitem [{\citenamefont {Ashhab}\ \emph {et~al.}(2008)\citenamefont {Ashhab},
  \citenamefont {Niskanen}, \citenamefont {Harrabi}, \citenamefont {Nakamura},
  \citenamefont {Picot}, \citenamefont {de~Groot}, \citenamefont {Harmans},
  \citenamefont {Mooij},\ and\ \citenamefont {Nori}}]{AshhabPRB08}%
  \BibitemOpen
  \bibfield  {author} {\bibinfo {author} {\bibfnamefont {S.}~\bibnamefont
  {Ashhab}}, \bibinfo {author} {\bibfnamefont {A.~O.}\ \bibnamefont
  {Niskanen}}, \bibinfo {author} {\bibfnamefont {K.}~\bibnamefont {Harrabi}},
  \bibinfo {author} {\bibfnamefont {Y.}~\bibnamefont {Nakamura}}, \bibinfo
  {author} {\bibfnamefont {T.}~\bibnamefont {Picot}}, \bibinfo {author}
  {\bibfnamefont {P.~C.}\ \bibnamefont {de~Groot}}, \bibinfo {author}
  {\bibfnamefont {C.~J. P.~M.}\ \bibnamefont {Harmans}}, \bibinfo {author}
  {\bibfnamefont {J.~E.}\ \bibnamefont {Mooij}}, \ and\ \bibinfo {author}
  {\bibfnamefont {F.}~\bibnamefont {Nori}},\ }\href@noop {} {\bibfield
  {journal} {\bibinfo  {journal} {Phys. Rev. B},\ }\textbf {\bibinfo {volume}
  {77}},\ \bibinfo {pages} {014510} (\bibinfo {year} {2008})}\BibitemShut
  {NoStop}%
\bibitem [{\citenamefont {Yamamoto}\ \emph {et~al.}(2008)\citenamefont
  {Yamamoto}, \citenamefont {Watanabe}, \citenamefont {You}, \citenamefont
  {Pashkin}, \citenamefont {Astafiev}, \citenamefont {Nakamura}, \citenamefont
  {Nori},\ and\ \citenamefont {Tsai}}]{YamamotoPRB08}%
  \BibitemOpen
  \bibfield  {author} {\bibinfo {author} {\bibfnamefont {T.}~\bibnamefont
  {Yamamoto}}, \bibinfo {author} {\bibfnamefont {M.}~\bibnamefont {Watanabe}},
  \bibinfo {author} {\bibfnamefont {J.~Q.}\ \bibnamefont {You}}, \bibinfo
  {author} {\bibfnamefont {Y.~A.}\ \bibnamefont {Pashkin}}, \bibinfo {author}
  {\bibfnamefont {O.}~\bibnamefont {Astafiev}}, \bibinfo {author}
  {\bibfnamefont {Y.}~\bibnamefont {Nakamura}}, \bibinfo {author}
  {\bibfnamefont {F.}~\bibnamefont {Nori}}, \ and\ \bibinfo {author}
  {\bibfnamefont {J.~S.}\ \bibnamefont {Tsai}},\ }\href@noop {} {\bibfield
  {journal} {\bibinfo  {journal} {Phys. Rev. B},\ }\textbf {\bibinfo {volume}
  {77}},\ \bibinfo {pages} {064505} (\bibinfo {year} {2008})}\BibitemShut
  {NoStop}%
\bibitem [{\citenamefont {Mariantoni}\ \emph {et~al.}(2008)\citenamefont
  {Mariantoni}, \citenamefont {Deppe}, \citenamefont {Marx}, \citenamefont
  {Gross}, \citenamefont {Wilhelm},\ and\ \citenamefont
  {Solano}}]{MariantoniPRB08}%
  \BibitemOpen
  \bibfield  {author} {\bibinfo {author} {\bibfnamefont {M.}~\bibnamefont
  {Mariantoni}}, \bibinfo {author} {\bibfnamefont {F.}~\bibnamefont {Deppe}},
  \bibinfo {author} {\bibfnamefont {A.}~\bibnamefont {Marx}}, \bibinfo {author}
  {\bibfnamefont {R.}~\bibnamefont {Gross}}, \bibinfo {author} {\bibfnamefont
  {F.~K.}\ \bibnamefont {Wilhelm}}, \ and\ \bibinfo {author} {\bibfnamefont
  {E.}~\bibnamefont {Solano}},\ }\href@noop {} {\bibfield  {journal} {\bibinfo
  {journal} {Phys. Rev. B},\ }\textbf {\bibinfo {volume} {78}},\ \bibinfo
  {pages} {104508} (\bibinfo {year} {2008})}\BibitemShut {NoStop}%
\bibitem [{\citenamefont {Allman}\ \emph {et~al.}(2010)\citenamefont {Allman},
  \citenamefont {Altomare}, \citenamefont {Whittaker}, \citenamefont {Cicak},
  \citenamefont {Li}, \citenamefont {Sirois}, \citenamefont {Strong},
  \citenamefont {Teufel},\ and\ \citenamefont {Simmonds}}]{AllmanPRL10}%
  \BibitemOpen
  \bibfield  {author} {\bibinfo {author} {\bibfnamefont {M.~S.}\ \bibnamefont
  {Allman}}, \bibinfo {author} {\bibfnamefont {F.}~\bibnamefont {Altomare}},
  \bibinfo {author} {\bibfnamefont {J.~D.}\ \bibnamefont {Whittaker}}, \bibinfo
  {author} {\bibfnamefont {K.}~\bibnamefont {Cicak}}, \bibinfo {author}
  {\bibfnamefont {D.}~\bibnamefont {Li}}, \bibinfo {author} {\bibfnamefont
  {A.}~\bibnamefont {Sirois}}, \bibinfo {author} {\bibfnamefont
  {J.}~\bibnamefont {Strong}}, \bibinfo {author} {\bibfnamefont {J.~D.}\
  \bibnamefont {Teufel}}, \ and\ \bibinfo {author} {\bibfnamefont {R.~W.}\
  \bibnamefont {Simmonds}},\ }\href@noop {} {\bibfield  {journal} {\bibinfo
  {journal} {Phys. Rev. Lett.},\ }\textbf {\bibinfo {volume} {104}},\ \bibinfo
  {pages} {177004} (\bibinfo {year} {2010})}\BibitemShut {NoStop}%
\bibitem [{\citenamefont {Pinto}\ \emph {et~al.}(2010)\citenamefont {Pinto},
  \citenamefont {Korotkov}, \citenamefont {Geller}, \citenamefont {Shumeiko},\
  and\ \citenamefont {Martinis}}]{PintoPRB10}%
  \BibitemOpen
  \bibfield  {author} {\bibinfo {author} {\bibfnamefont {R.~A.}\ \bibnamefont
  {Pinto}}, \bibinfo {author} {\bibfnamefont {A.~N.}\ \bibnamefont {Korotkov}},
  \bibinfo {author} {\bibfnamefont {M.~R.}\ \bibnamefont {Geller}}, \bibinfo
  {author} {\bibfnamefont {V.~S.}\ \bibnamefont {Shumeiko}}, \ and\ \bibinfo
  {author} {\bibfnamefont {J.~M.}\ \bibnamefont {Martinis}},\ }\href@noop {}
  {\bibfield  {journal} {\bibinfo  {journal} {Phys. Rev. B},\ }\textbf
  {\bibinfo {volume} {83}},\ \bibinfo {pages} {104522} (\bibinfo {year}
  {2010})}\BibitemShut {NoStop}%
\bibitem [{\citenamefont {Gambetta}\ \emph {et~al.}(2011)\citenamefont
  {Gambetta}, \citenamefont {Houck},\ and\ \citenamefont
  {Blais}}]{GambettaPRL11}%
  \BibitemOpen
  \bibfield  {author} {\bibinfo {author} {\bibfnamefont {J.~M.}\ \bibnamefont
  {Gambetta}}, \bibinfo {author} {\bibfnamefont {A.~A.}\ \bibnamefont {Houck}},
  \ and\ \bibinfo {author} {\bibfnamefont {A.}~\bibnamefont {Blais}},\
  }\href@noop {} {\bibfield  {journal} {\bibinfo  {journal} {Phys. Rev.
  Lett.},\ }\textbf {\bibinfo {volume} {106}},\ \bibinfo {pages} {030502}
  (\bibinfo {year} {2011})}\BibitemShut {NoStop}%
\bibitem [{\citenamefont {Bialczak}\ \emph {et~al.}(2011)\citenamefont
  {Bialczak}, \citenamefont {Ansmann}, \citenamefont {Hofheinz}, \citenamefont
  {Lenander}, \citenamefont {Lucero}, \citenamefont {Neeley}, \citenamefont
  {O'Connell}, \citenamefont {Sank}, \citenamefont {Wang}, \citenamefont
  {Weides}, \citenamefont {Wenner}, \citenamefont {Yamamoto}, \citenamefont
  {Cleland},\ and\ \citenamefont {Martinis}}]{BialczakPRL11}%
  \BibitemOpen
  \bibfield  {author} {\bibinfo {author} {\bibfnamefont {R.~C.}\ \bibnamefont
  {Bialczak}}, \bibinfo {author} {\bibfnamefont {M.}~\bibnamefont {Ansmann}},
  \bibinfo {author} {\bibfnamefont {M.}~\bibnamefont {Hofheinz}}, \bibinfo
  {author} {\bibfnamefont {M.}~\bibnamefont {Lenander}}, \bibinfo {author}
  {\bibfnamefont {E.}~\bibnamefont {Lucero}}, \bibinfo {author} {\bibfnamefont
  {M.}~\bibnamefont {Neeley}}, \bibinfo {author} {\bibfnamefont {A.~D.}\
  \bibnamefont {O'Connell}}, \bibinfo {author} {\bibfnamefont {D.}~\bibnamefont
  {Sank}}, \bibinfo {author} {\bibfnamefont {H.}~\bibnamefont {Wang}}, \bibinfo
  {author} {\bibfnamefont {M.}~\bibnamefont {Weides}}, \bibinfo {author}
  {\bibfnamefont {J.}~\bibnamefont {Wenner}}, \bibinfo {author} {\bibfnamefont
  {T.}~\bibnamefont {Yamamoto}}, \bibinfo {author} {\bibfnamefont {A.~N.}\
  \bibnamefont {Cleland}}, \ and\ \bibinfo {author} {\bibfnamefont {J.~M.}\
  \bibnamefont {Martinis}},\ }\href@noop {} {\bibfield  {journal} {\bibinfo
  {journal} {Phys. Rev. Lett.},\ }\textbf {\bibinfo {volume} {106}},\ \bibinfo
  {pages} {060501} (\bibinfo {year} {2011})}\BibitemShut {NoStop}%
\bibitem [{\citenamefont {Srinivasan}\ \emph {et~al.}(2011)\citenamefont
  {Srinivasan}, \citenamefont {Hoffman}, \citenamefont {Gambetta},\ and\
  \citenamefont {Houck}}]{SrinivasanPRL11}%
  \BibitemOpen
  \bibfield  {author} {\bibinfo {author} {\bibfnamefont {S.~J.}\ \bibnamefont
  {Srinivasan}}, \bibinfo {author} {\bibfnamefont {A.~J.}\ \bibnamefont
  {Hoffman}}, \bibinfo {author} {\bibfnamefont {J.~M.}\ \bibnamefont
  {Gambetta}}, \ and\ \bibinfo {author} {\bibfnamefont {A.~A.}\ \bibnamefont
  {Houck}},\ }\href@noop {} {\bibfield  {journal} {\bibinfo  {journal} {Phys.
  Rev. Lett.},\ }\textbf {\bibinfo {volume} {106}},\ \bibinfo {pages} {083601}
  (\bibinfo {year} {2011})}\BibitemShut {NoStop}%
\bibitem [{\citenamefont {Groszkowski}\ \emph {et~al.}(2011)\citenamefont
  {Groszkowski}, \citenamefont {Fowler}, \citenamefont {Motzoi},\ and\
  \citenamefont {Wilhelm}}]{GroszkowskiPRB11}%
  \BibitemOpen
  \bibfield  {author} {\bibinfo {author} {\bibfnamefont {P.}~\bibnamefont
  {Groszkowski}}, \bibinfo {author} {\bibfnamefont {A.~G.}\ \bibnamefont
  {Fowler}}, \bibinfo {author} {\bibfnamefont {F.}~\bibnamefont {Motzoi}}, \
  and\ \bibinfo {author} {\bibfnamefont {F.~K.}\ \bibnamefont {Wilhelm}},\
  }\href@noop {} {\bibfield  {journal} {\bibinfo  {journal} {Phys. Rev. B},\
  }\textbf {\bibinfo {volume} {84}},\ \bibinfo {pages} {144516} (\bibinfo
  {year} {2011})}\BibitemShut {NoStop}%
\bibitem [{\citenamefont {Hoffman}\ \emph {et~al.}(2011)\citenamefont
  {Hoffman}, \citenamefont {Srinivasan}, \citenamefont {Gambetta},\ and\
  \citenamefont {Houck}}]{HoffmanPRB11}%
  \BibitemOpen
  \bibfield  {author} {\bibinfo {author} {\bibfnamefont {A.~J.}\ \bibnamefont
  {Hoffman}}, \bibinfo {author} {\bibfnamefont {S.~J.}\ \bibnamefont
  {Srinivasan}}, \bibinfo {author} {\bibfnamefont {J.~M.}\ \bibnamefont
  {Gambetta}}, \ and\ \bibinfo {author} {\bibfnamefont {A.~A.}\ \bibnamefont
  {Houck}},\ }\href@noop {} {\bibfield  {journal} {\bibinfo  {journal} {Phys.
  Rev. B},\ }\textbf {\bibinfo {volume} {84}},\ \bibinfo {pages} {184515}
  (\bibinfo {year} {2011})}\BibitemShut {NoStop}%
\bibitem [{\citenamefont {Koch}\ \emph {et~al.}(2007)\citenamefont {Koch},
  \citenamefont {Yu}, \citenamefont {Gambetta}, \citenamefont {Houck},
  \citenamefont {Schuster}, \citenamefont {Majer}, \citenamefont {Blais},
  \citenamefont {Devoret}, \citenamefont {Girvin},\ and\ \citenamefont
  {Schoelkopf}}]{KochPRA07}%
  \BibitemOpen
  \bibfield  {author} {\bibinfo {author} {\bibfnamefont {J.}~\bibnamefont
  {Koch}}, \bibinfo {author} {\bibfnamefont {T.~M.}\ \bibnamefont {Yu}},
  \bibinfo {author} {\bibfnamefont {J.}~\bibnamefont {Gambetta}}, \bibinfo
  {author} {\bibfnamefont {A.~A.}\ \bibnamefont {Houck}}, \bibinfo {author}
  {\bibfnamefont {D.~I.}\ \bibnamefont {Schuster}}, \bibinfo {author}
  {\bibfnamefont {J.}~\bibnamefont {Majer}}, \bibinfo {author} {\bibfnamefont
  {A.}~\bibnamefont {Blais}}, \bibinfo {author} {\bibfnamefont {M.~H.}\
  \bibnamefont {Devoret}}, \bibinfo {author} {\bibfnamefont {S.~M.}\
  \bibnamefont {Girvin}}, \ and\ \bibinfo {author} {\bibfnamefont {R.~J.}\
  \bibnamefont {Schoelkopf}},\ }\href@noop {} {\bibfield  {journal} {\bibinfo
  {journal} {Phys. Rev. A},\ }\textbf {\bibinfo {volume} {76}},\ \bibinfo
  {pages} {042319} (\bibinfo {year} {2007})}\BibitemShut {NoStop}%
\bibitem [{\citenamefont {Chen}\ \emph {et~al.}()\citenamefont {Chen},
  \citenamefont {Neill}, \citenamefont {Roushan}, \citenamefont {Leung},
  \citenamefont {Fang}, \citenamefont {Barends}, \citenamefont {Kelly},
  \citenamefont {Campbell}, \citenamefont {Chen}, \citenamefont {Chiaro},
  \citenamefont {Dunsworth}, \citenamefont {Jeffrey}, \citenamefont {Megrant},
  \citenamefont {Mutus}, \citenamefont {O$^\prime$Malley}, \citenamefont
  {Quintana}, \citenamefont {Sank}, \citenamefont {Vainsencher}, \citenamefont
  {Wenner}, \citenamefont {White}, \citenamefont {Geller}, \citenamefont
  {Cleland},\ and\ \citenamefont {Martinis}}]{ChenEtal1402.7367}%
  \BibitemOpen
  \bibfield  {author} {\bibinfo {author} {\bibfnamefont {Y.}~\bibnamefont
  {Chen}}, \bibinfo {author} {\bibfnamefont {C.}~\bibnamefont {Neill}},
  \bibinfo {author} {\bibfnamefont {P.}~\bibnamefont {Roushan}}, \bibinfo
  {author} {\bibfnamefont {N.}~\bibnamefont {Leung}}, \bibinfo {author}
  {\bibfnamefont {M.}~\bibnamefont {Fang}}, \bibinfo {author} {\bibfnamefont
  {R.}~\bibnamefont {Barends}}, \bibinfo {author} {\bibfnamefont
  {J.}~\bibnamefont {Kelly}}, \bibinfo {author} {\bibfnamefont
  {B.}~\bibnamefont {Campbell}}, \bibinfo {author} {\bibfnamefont
  {Z.}~\bibnamefont {Chen}}, \bibinfo {author} {\bibfnamefont {B.}~\bibnamefont
  {Chiaro}}, \bibinfo {author} {\bibfnamefont {A.}~\bibnamefont {Dunsworth}},
  \bibinfo {author} {\bibfnamefont {E.}~\bibnamefont {Jeffrey}}, \bibinfo
  {author} {\bibfnamefont {A.}~\bibnamefont {Megrant}}, \bibinfo {author}
  {\bibfnamefont {J.~Y.}\ \bibnamefont {Mutus}}, \bibinfo {author}
  {\bibfnamefont {P.~J.~J.}\ \bibnamefont {O$^\prime$Malley}}, \bibinfo
  {author} {\bibfnamefont {C.~M.}\ \bibnamefont {Quintana}}, \bibinfo {author}
  {\bibfnamefont {D.}~\bibnamefont {Sank}}, \bibinfo {author} {\bibfnamefont
  {A.}~\bibnamefont {Vainsencher}}, \bibinfo {author} {\bibfnamefont
  {J.}~\bibnamefont {Wenner}}, \bibinfo {author} {\bibfnamefont {T.~C.}\
  \bibnamefont {White}}, \bibinfo {author} {\bibfnamefont {M.~R.}\ \bibnamefont
  {Geller}}, \bibinfo {author} {\bibfnamefont {A.~N.}\ \bibnamefont {Cleland}},
  \ and\ \bibinfo {author} {\bibfnamefont {J.~M.}\ \bibnamefont {Martinis}},\
  }\href@noop {} {\enquote {\bibinfo {title} {Qubit architecture with high
  coherence and fast tunable coupling},}\ }\bibinfo {note}
  {{a}rXiv:1402.7367}\BibitemShut {NoStop}%
\end{thebibliography}%

\end{document}